\documentclass{aa}  
\graphicspath{{./}{figures/}}
\bibpunct{(}{)}{;}{a}{}{,} 
\usepackage{graphicx, xcolor}
\usepackage{txfonts}
\usepackage{placeins}
\usepackage{hyperref}
\hypersetup{
    colorlinks=True,
    allcolors=blue
    }
\makeatletter
\renewcommand*\aa@pageof{, page \thepage{} of \pageref*{LastPage}}
\makeatother

\begin{document}

   \title{SRG/eROSITA X-ray shadowing study of giant molecular clouds}

   \author{M.~C.~H.~Yeung
          \inst{1}
          \and
          M.~J.~Freyberg\inst{1}
          \and 
          G.~Ponti \inst{2}\fnmsep\inst{1}
          \and
          K.~Dennerl\inst{1}
          \and
          M.~Sasaki\inst{3}
          \and
          A.~Strong\inst{1}}

   \institute{Max-Planck-Institut für extraterrestrische Physik, Giessenbachstraße, 85748 Garching, Germany\\
              \email{myeung@mpe.mpg.de}
         \and
             INAF-Osservatorio Astronomico di Brera, Via E. Bianchi 46, I-23807 Merate (LC), Italy
         \and
             Dr.~Karl Remeis Observatory, Erlangen Centre for Astroparticle Physics, Friedrich-Alexander-Universit{\"a}t Erlangen-N{\"u}rnberg, Sternwartstra{\ss}e 7, 96049 Bamberg, Germany}
   \authorrunning{M.~C.~H.~Yeung et al.}

 
  \abstract
   {SRG/eROSITA is situated in a halo orbit around L2 where the highly variable solar wind charge exchange (SWCX) emission from Earth's magnetosheath is 
   expected to be negligible. 
   The soft X-ray foreground emissions from the local hot bubble (LHB) and the remaining heliospheric SWCX emissions could be studied in unprecedented detail with eROSITA All-Sky Survey (eRASS) data in a 6-month cadence and better spectral resolution than ROSAT.
   }
   {We aim to use eRASS data of the sight lines towards three giant molecular clouds away from the Galactic plane to isolate and study the soft X-ray diffuse foreground emission. These X-ray shadows will serve as calibration baselines for the future three-dimensional structural study of the LHB.}
   {We conducted spectral analysis on the diffuse X-ray spectra of these clouds from the first four eRASSs to estimate and separate the heliospheric SWCX contribution from the LHB emission.}
   {We find the density of the LHB to be independent of the sight line with $n_e \sim 4 \times 10^{-3}$\,cm$^{-3}$, but not the temperature. We report a lower temperature of $kT_{\mathrm{LHB}}=0.084\pm0.004$\,keV towards Chamaeleon~II \& III (Cha~II \& III) than Ophiuchus (Oph) and Corona Australis (CrA), in which we measured $0.102\pm0.006$ and $0.112\pm0.009$\,keV, respectively. 
   We measured the emission measure of the LHB to be $\sim 2\times10^{-3}\,$cm$^{-6}$\,pc at medium Galactic latitudes ($|b|\sim 20\degr$).
   A monotonic increase in the SWCX contribution has been observed since the start of 2020, coincidental with the beginning of solar cycle 25. For Oph, SWCX has dominated the LHB in the $0.3$--$0.7$\,keV band intensity since eRASS2. We observed lower SWCX contributions in Cha~II \& III and CrA, consistent with the expected decreasing solar wind ion density at high heliographic latitudes.}
   {}

   \date{Received 9 January 2023 / Accepted 18 April 2023}
   \maketitle
   
%
\section{Introduction}

The notion of the soft X-ray emitting local hot bubble (LHB) emerged after Wisconsin sounding-rocket data showed an anti-correlation between the soft X-ray intensity and neutral hydrogen column density ($N_{\element{H}}$) in the southern Galactic hemisphere \citep{Sanders1977}. In their concise yet seminal paper, \citet{Sanders1977} pointed out that the anti-correlation could not be accommodated for by photoelectric absorption but by a displacement effect in which the local volume is filled with an X-ray emitting gas bounded by a thick wall of cool neutral hydrogen gas. In this picture, a low $N_{\element{H}}$ (thin) section of the wall is displaced by additional X-ray emitting gas, resulting in a higher X-ray intensity in low $N_{\element{H}}$ regions. Independently, \citet{Tanaker_Bleeker_1977} reached the same conclusion and coined the term LHB.
The existence of the LHB had henceforth become the standard conceptual picture of the local interstellar medium (ISM) --- a low \ion{H}{I} column density region ($N_{\element{H}} \lesssim 10^{20}$~cm$^{-2}$) filled with hot plasma extending to $100$--$200$~pc \citep[e.g. see reviews by][]{Bochkarev_review,McCammon_Sanders_review,Breitschwerdt_review}. After the launch of ROSAT, shadowing experiments on Draco and MBM12 clouds firmly indicated the presence of a more distant soft X-ray background from the Galaxy in addition to the foreground LHB emission \citep[e.g.][]{Snowden_1991,Snowden_MBM12}.

However, ROSAT delivered arguably even more insights into the additional soft emission, which turned out to be foreground emission and led to the discovery of the solar wind charge exchange (SWCX) process. It began with the detection of long-term enhancements (LTEs) in the ROSAT All-Sky Survey (RASS) with durations ranging up to $\sim 8$~hours, which were then found to be correlated with solar wind variations and geomagnetic storms \citep{Freyberg_thesis}. Later detection of an unexpectedly bright soft X-ray from the Comet C/Hyakutake 1996 B2 with an emission morphology facing the Sun but not the direction of motion strongly suggests that the Sun is the culprit \citep{Lisse_1996}. Soon after, \citet{Cravens1997} proposed that the charge exchange process between the cometary neutrals and heavy solar wind ions could explain the comet's soft X-ray emission.
\citet{Konrad1997} established comets as a class of X-ray sources by systematically searching the archival ROSAT data and they pointed out that the charge exchange between highly charged solar wind ions and cometary neutrals is the dominant emission process.
A LTE in the X-ray background was also detected near the outbursts of the comet, prompting \citet{Konrad1997} and \citet{Freyberg1998} to suggest the Earth could also act as a bright, soft X-ray source from the SWCX process. Concurrently, \citet{Cox1998} pointed out the flowing neutral ISM also provides a source of neutrals to interact with the solar wind. The SWCX emissions from these two sources of neutral atoms are usually referred to as magnetospheric and heliospheric SWCX, respectively \citep[see review by][]{Kuntz2019}.

Distinguishing between the contributions of LHB and SWCX using ROSAT PSPC is challenging due to the reliance on broad-band count rates. Many cross sections of the heavy ions were unknown, further complicating the issue. Before the arrival of more SWCX-focussed missions such as the Diffuse X-rays from the Local Galaxy (DXL), simultaneous modelling of the LHB and SWCX have resulted in sometimes inconsistent results --- from estimations of SWCX contributing to half to all of the $\frac{1}{4}$~keV emission in the galactic plane \citep[][and references therein]{Koutroumpa2009,Robertson2009}. There was naturally a worry that LHB had become redundant and all the $\frac{1}{4}$~keV emission could be accounted for by SWCX, despite a consistent requirement of additional emission at high galactic latitudes also shown by these studies.

\citet{Galeazzi2014} mostly settled the situation by estimating the heliospheric SWCX contribution in ROSAT using the DXL sounding rocket mission and reported that the LHB still contributes to $\sim 60$\% of the emission in the galactic plane. Based on the estimated SWCX contribution from DXL \citep{DXL,Uprety}, \citet{Liu16} measured a LHB temperature of $0.097\pm0.019$~keV from the ROSAT R2/R1 band ratio map and mapped out the three-dimensional (3D) structure of the LHB assuming a constant electron density plasma. They show that the inferred boundary of the LHB agrees reasonably well with the onset of a higher absorbing column inferred from the local ISM density map \citep{Lallement2014}. 

With energy-resolved imaging and repeated all-sky surveys every six months, eROSITA is providing an unprecedented view of the soft X-ray foreground in terms of the depth and differentiation of LHB from SWCX. Situated in a halo orbit around L2, eROSITA is expected to be free of the highly variable SWCX coming from the Earth's magnetosheath. The energy resolution of $\Delta E \simeq58$~eV at the C-K line ($0.277$~keV) allows for spectral decomposition of the heliospheric SWCX and LHB components \citep{eROSITA}, as well as measurements of the properties of the LHB and SWCX intensity in six-month cadence along the chosen sight lines.

In this paper, we study the emissions from three of the darkest X-ray shadows away from the Galactic plane in the German eROSITA sky (Galactic coordinates restricted to $180\degr \le l \le 360\degr$, $-90\degr \le b \le 90\degr$) --- Chamaeleon~II \& III (Cha~II \& III), Ophiuchus (Oph), and Corona Australis (CrA). They lie on the boundary of the LHB \citep{zucker} with accurate distances and are, therefore, ideal calibration points of the LHB properties, which could later be extended to infer the 3D structure of the LHB. Earlier observational work based on ROSAT found the LHB is well described by a single temperature and density plasma \citep[e.g.][]{Liu16}. We aim to subject this assumption to the tighter constraints set by eROSITA. These sight lines also enable us to infer the time evolution of the relative contribution between the LHB and heliospheric SWCX.

In Sect.\,\ref{sec:data}, we describe the eROSITA observations of the clouds and the specific regions chosen for spectral extraction, as well as the XMM-Newton celestial calibration source as an independent monitoring of the SWCX variation. We lay out the constituents of our spectral model and our spectral fitting procedures in Sect.\,\ref{sec:ana}. We report our results and interpretations in Sect.\,\ref{sec:result}. Finally, we deliver our concluding remarks in Sect.\,\ref{sec:con}.

\section{Data and calibration} \label{sec:data}
\begin{figure*} 
    \centering
    \includegraphics[width=\textwidth]{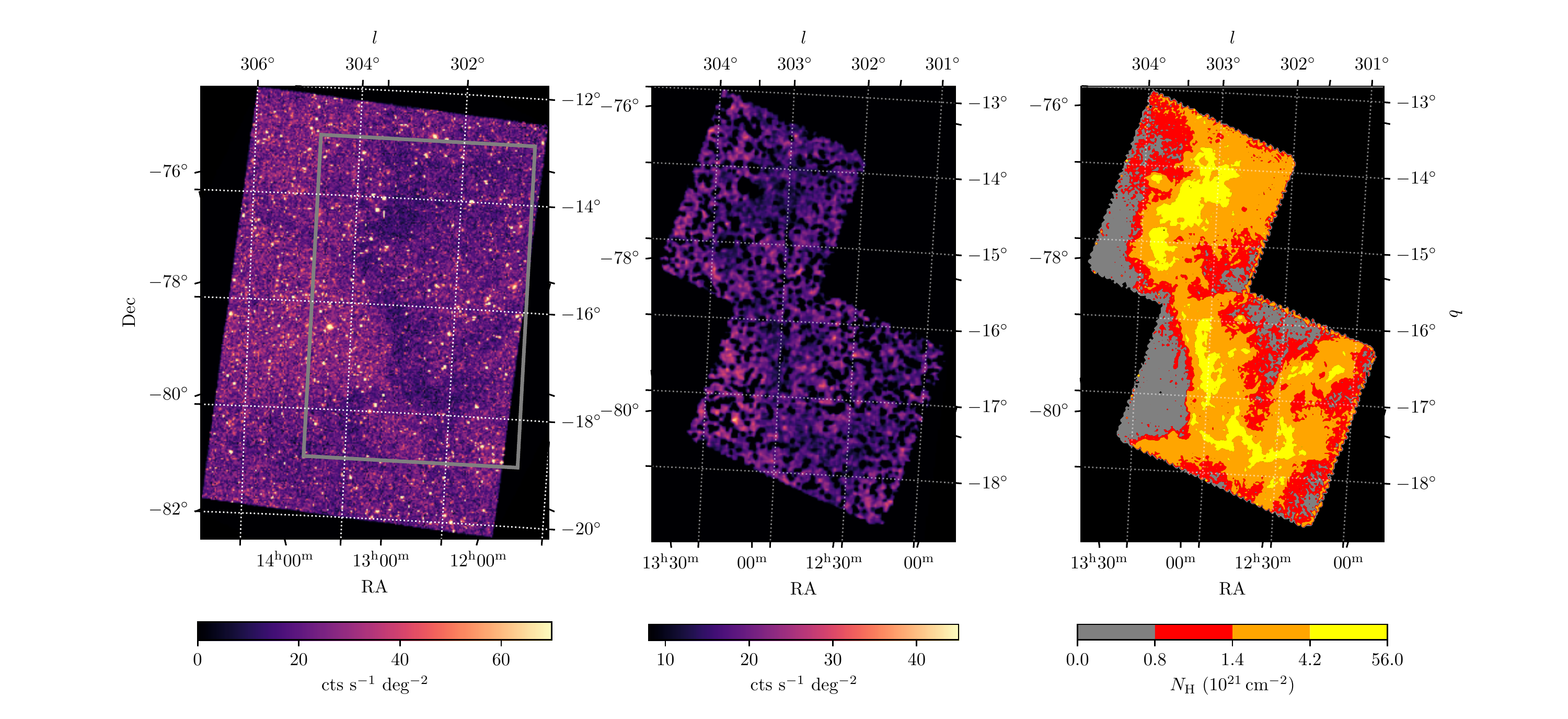}
    \caption{eROSITA 0.2--3\,keV image of the shadow cast by Cha~II \& III. The left panel shows the eROSITA 0.2--3~keV band image of Cha~II \& III colour-coded with the vignetting-corrected count rate. The grey rectangular box indicates the region shown in the middle and right plots. The middle panel shows the point-source-free region where spectral analysis is carried out. The two rectangular mosaics represent the region covered by the \textit{Herschel} column density map. The right panel shows the four column density bins that define the regions for spectral extraction. The dotted grid is in the Galactic coordinate system.}
    \label{fig:img}
\end{figure*}

\begin{figure*} 
    \centering
    \includegraphics[width=\textwidth]{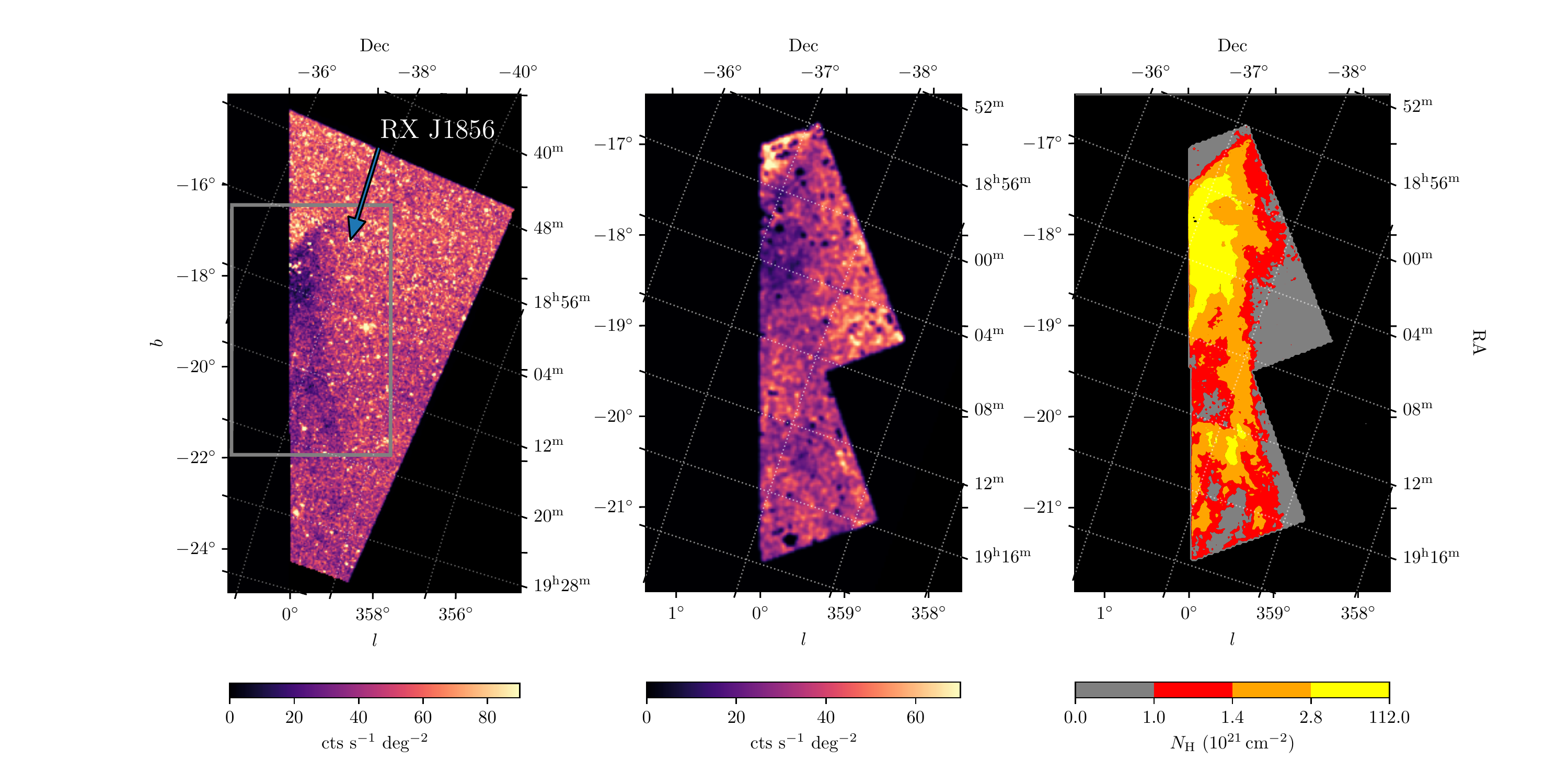}
    \caption{Same as Fig.\,\ref{fig:img}, but for CrA. The dotted grid is in the equatorial coordinate system.}
    \label{fig:rcra_img}
\end{figure*}

\begin{figure*} 
    \centering
    \includegraphics[width=\textwidth]{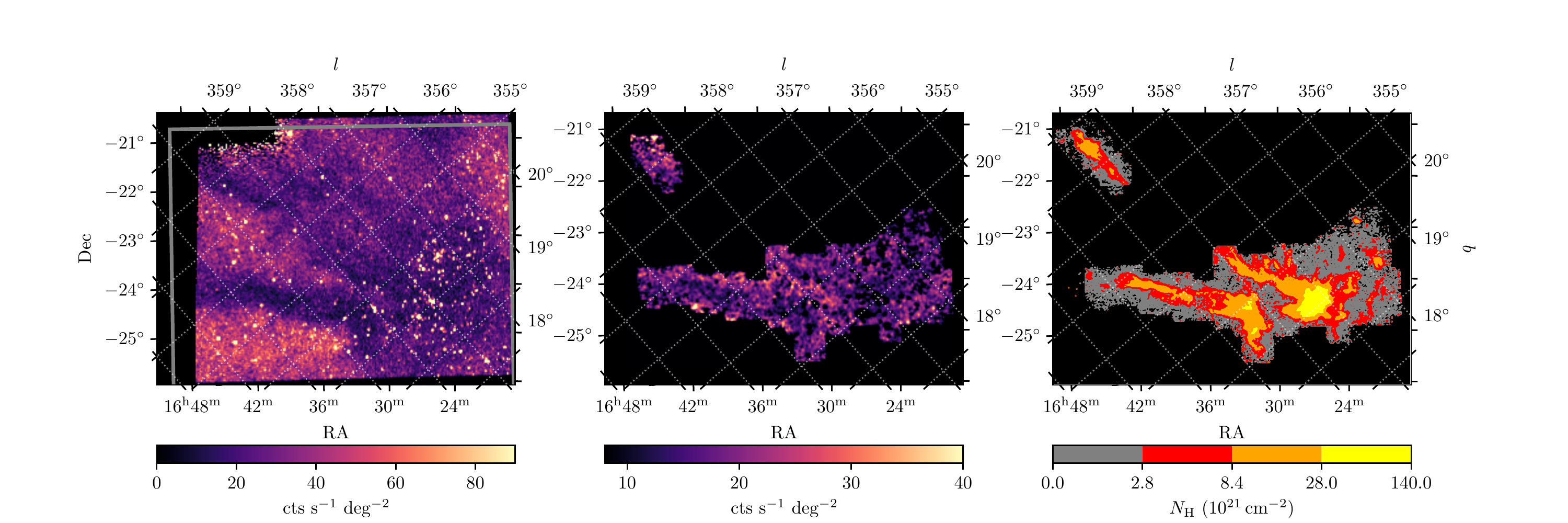}
    \caption{Same as Fig.\,\ref{fig:img}, but for Oph. The dotted grid is in the Galactic coordinate system, with $l$ increasing towards the top left and $b$ increasing towards the top right.}
    \label{fig:oph_img}
\end{figure*}

We extracted data of Cha~II \& III, Oph and CrA from the first four eROSITA All-Sky Surveys (eRASSs). Only data from telescope modules (TM) with on-chip filters (TM 1--4 and 6) were extracted to avoid the light leak issue \citep{eROSITA}. Four visits to each cloud enable an analysis of the variability over intervals of half a year. The data were processed with the 020 version of the eSASS pipeline \citep{Brunner2022}. Flares were removed using the standard eSASS task \texttt{flaregti}. After removing the flared time intervals, the total vignetting-corrected exposure times within the defined regions (right panel of Figs.\,\ref{fig:img}, \ref{fig:rcra_img} and \ref{fig:oph_img}) in the first four eRASSs in the $0.2$--$3$\,keV band (our spectral-fitting range) are $\sim 760$~s for Cha~II \& III, $\sim 330$~s for Oph, and $\sim 250$~s for CrA. The difference is mainly caused by their angular distance from the ecliptic poles, where all the great circles of the eROSITA scans merge to result in the maximum exposure time. 

All valid event patterns were used because the low-energy electronic noise component (mostly $\lesssim 0.3$~keV) has been greatly reduced from the 946 (as used in the Early Data Release \citep{Brunner2022}) to 020 processing version. Thus one could take advantage of more photons to constrain the spectral models at low energies and need not sacrifice higher-pattern events to suppress the electronic noise.

The 0.2--3~keV images of the clouds are shown in the left panels of Figs.\,\ref{fig:img}, \ref{fig:rcra_img} and \ref{fig:oph_img}. One can observe a clear shadow between $301 \degr \lesssim l \lesssim 304\degr$ and $13 \degr \lesssim b \lesssim 18 \degr$ cast by Cha~II \& III. The X-ray shadows are even more prominent in the case of CrA and Oph --- a consequence of the two clouds being in front of and absorbing the emissions from the bright eROSITA bubbles \citep{eROSITA_bubble}. Point sources were masked using the CheeseMask images produced by the standard eSASS detection chain, as shown in the middle panels.

In our spectral analysis, spectra of varying column density regions were extracted. The extraction regions are colour-coded on the right panels of Figs.\,\ref{fig:img}, \ref{fig:rcra_img} and \ref{fig:oph_img} by column density. The hydrogen column density information was obtained from the \textit{Herschel} Gould Belt Survey Archive \citep{Herschel}. The column density maps\footnote{The \textit{Herschel} $N_{\mathrm{H_2}}$ maps were converted to $N_{\mathrm{H}}$ using $\frac{\mu_{\mathrm{H_2}}}{\mu_{\mathrm{H}}}=\frac{2.8}{1.37}\simeq2$ \citep{Roy2014}} were produced by fitting the SEDs formed by the 160, 250, 350 and 500~$\mu$m images in a 6\arcsec grid at an angular resolution of 18\farcs2 \citep{cha_herschel,Bresnahan2018,oph_herschel}. We convolved the column density map to the approximate eROSITA angular resolution of $30$\arcsec, then defined four regions bound by the contour levels. Regions of Cha~II \& III and CrA are defined in this way. Table~\ref{tab:exposure} lists the solid angles of the defined regions.

    

\begin{table*}
    \centering
    \begin{tabular}{ccccccc}
    \hline\hline
    Cloud & Exposure Time (s) & Region    & Area (deg$^2$) & \multicolumn{3}{c}{$N_\mathrm{H}$ ($10^{21}$\,cm$^{-2}$)}\\
          && &   & \textit{Herschel}  & \textit{Planck} $R$ & \textit{Planck} $\tau_{353}$ \\\hline
 && 1 & 1.69& $0.47^{+0.11}_{-0.12}$ & $0.75^{+0.14}_{-0.15}$ &  $0.84^{+0.16}_{-0.19}$  \\ 

 Cha~II \& III & 762 &2 &2.21& $0.80^{+0.14}_{-0.12}$ & $1.01^{+0.11}_{-0.13}$ &  $1.31^{+0.22}_{-0.18}$  \\ 

 && 3 &4.14 &$1.55^{+0.52}_{-0.33}$ & $1.43^{+0.27}_{-0.17}$ &  $2.36^{+0.67}_{-0.39}$  \\ 

 && 4 & 1.15&$4.10^{+1.02}_{-0.73}$ & $2.34^{+0.38}_{-0.28}$ &  $4.86^{+1.19}_{-0.87}$  \\ 
  \hline

  && 1 & 1.37 & $0.55^{+0.10}_{-0.09}$ & $0.60^{+0.11}_{-0.09}$ &  $0.60^{+0.21}_{-0.15}$  \\ 
  
 CrA & 251 &2 & 1.40 & $0.86^{+0.07}_{-0.08}$ & $0.89^{+0.12}_{-0.11}$ &  $1.08^{+0.17}_{-0.19}$  \\ 

 && 3 & 1.32 & $1.25^{+0.24}_{-0.14}$ & $1.17^{+0.26}_{-0.11}$ &  $1.64^{+0.34}_{-0.23}$  \\

 && 4 &0.60& $3.54^{+3.09}_{-1.01}$ & $2.58^{+1.07}_{-0.55}$ &  $4.56^{+2.33}_{-1.10}$  \\ 
 \hline
 && 1 & 3.77&$0.93^{+0.52}_{-0.37}$ & $7.86^{+4.66}_{-3.61}$ &  $4.35^{+1.02}_{-0.77}$  \\ 

 Oph & 330 &2 &2.49 & $3.24^{+1.12}_{-1.18}$ & $8.69^{+4.41}_{-3.21}$ &  $5.73^{+1.33}_{-1.05}$  \\ 

 && 3 &1.40& $8.43^{+2.96}_{-1.90}$ & $11.07^{+4.77}_{-3.68}$ &  $8.41^{+2.44}_{-1.22}$  \\ 

 && 4 &0.19& $27.89^{+6.56}_{-5.29}$ & $38.90^{+20.16}_{-13.44}$ &  $17.91^{+3.01}_{-3.46}$  \\  \hline
    
    \end{tabular}
    \caption{Summary of the exposure times and the extraction regions in each molecular cloud.}
    \tablefoot{The exposure time is the average vignetting-corrected on-axis exposure time, assuming a nominal seven-TM effective area in the $0.2$--$3$\,keV band. The $N_\mathrm{H}$ values from \textit{Herschel} and \textit{Planck} are the $50^{\mathrm{th}}$ percentile in each extraction region after convolving the maps to a common angular resolution of 5\arcmin. with the lower and upper bounds showing the $25^{\mathrm{th}}$ and $75^{\mathrm{th}}$ percentiles respectively.}
    \label{tab:exposure}
\end{table*}

For Ophiuchus, the area coverage of \textit{Herschel}'s column density map is limited to the cloud core. We extended the area with column density information by using the \element[][13]{C}\element{O}~1--0 (110.201~GHz) map observed by the 14~m Five College Radio Astronomy Observatory (FCRAO) telescope \citep{Ridge06}. This was done by deriving the mean \element[][13]{C}\element{O}-to-$N_{\mathrm{H}}$ conversion factor from the overlap region of the \textit{Herschel} and FCRAO maps while taking into account the difference in angular resolution and hence applying this factor to the additional area that the \element[][13]{CO} map possesses. The result, in the form of the four regions' contours, is shown in the right panel of Fig.\,\ref{fig:oph_img}.

While the \textit{Herschel} column density maps provide the exceptional angular resolution necessary for this work, similar to the \textit{Planck} radiance map ($R$) that also adopts SED fitting to extract $N_\mathrm{H}$ information, they are likely to be affected by variations in the radiation field strength caused by increased attenuation of the interstellar radiation field and local heating photons in molecular clouds \citep{planck2014}. \citet{planck2014} suggests the dust opacity at 353\,GHz ($\tau_{353}$) is a better tracer of $N_{\mathrm{H}}$ in these regions. To estimate the possible range of $N_{\mathrm{H}}$ in each extraction region, we computed the $N_{\mathrm{H}}$ inferred from the aforementioned tracers and list the results in Table\,\ref{tab:exposure}.

All the maps were convolved to the common angular resolution of 5\arcmin (resolution of the \textit{Planck} maps) to ensure a fair comparison. The \textit{Planck} maps were first converted to the corresponding $E(B-V)$ maps using the relations $E(B-V)/R=(5.40\pm0.09)\times10^5$ and $E(B-V)/\tau_{353}=(1.49\pm0.03)\times10^4$ using quasars in the diffuse ISM at high Galactic latitudes \citep{planck2014}. Subsequently, we adopted the scaling of $N_\mathrm{H}/E(B-V)=4\times10^{21}$\,cm$^{-2}$\,mag$^{-1}$ to get $N_{\mathrm{H}}$ from the $E(B-V)$ maps. This scaling is representative of values derived from the detailed multiphase analysis conducted by \citet{PlanckFermi} and is shown by \citet{Lallement2016} to match the fitted X-ray foreground absorption in 19 XMM-Newton sight lines towards the North Polar Spur.

We note the reasonable agreement between the three tracers, given that there is still a large scatter with each scaling relation we used to convert $R$ and $\tau_{353}$ to $N_{\mathrm{H}}$. The variations of $N_{\mathrm{H}}$ inferred from different tracers inform the scale of systematic uncertainties in our derivation of 
 $N_{\mathrm{H}}$, which is instructive to compare with the best-fit $N_\mathrm{H}$ values in Table\,\ref{table:best-fit} in Sect.\,\ref{sec:result}. Interestingly, a direct comparison between our best-fit $N_\mathrm{H}$ and the values derived from the three tracers would suggest $\tau_{353}$ might be a better $N_\mathrm{H}$ tracer towards these high $N_\mathrm{H}$ regions, despite the fact that the region boundaries are drawn from the \textit{Herschel} maps. 
An exception to the overall agreement is found in region 1 of Oph. The reason for this is likely to be the inaccurate extrapolation of the \textit{Herschel} map using $^{13}$CO map when the latter approaches the sensitivity limit in the lowest $N_\mathrm{H}$ region.
 
Background count rates in the $0.3$--$0.7$\,keV and $8$-$12$\,keV bands of the XMM-Newton routine calibration source RX~J1856.5-3754 (RX~J1856) were also extracted in order to compare and cross-check our findings on SWCX towards the direction of CrA. The position of RX~J1856 relative to CrA is indicated on the left panel of Fig.\,\ref{fig:rcra_img}. Appendix~\ref{appendix:rxj} describes the datasets and the results in more detail.

\section{Spectral Analysis} \label{sec:ana}
\begin{figure}
    \centering
    \includegraphics[width=0.45\textwidth]{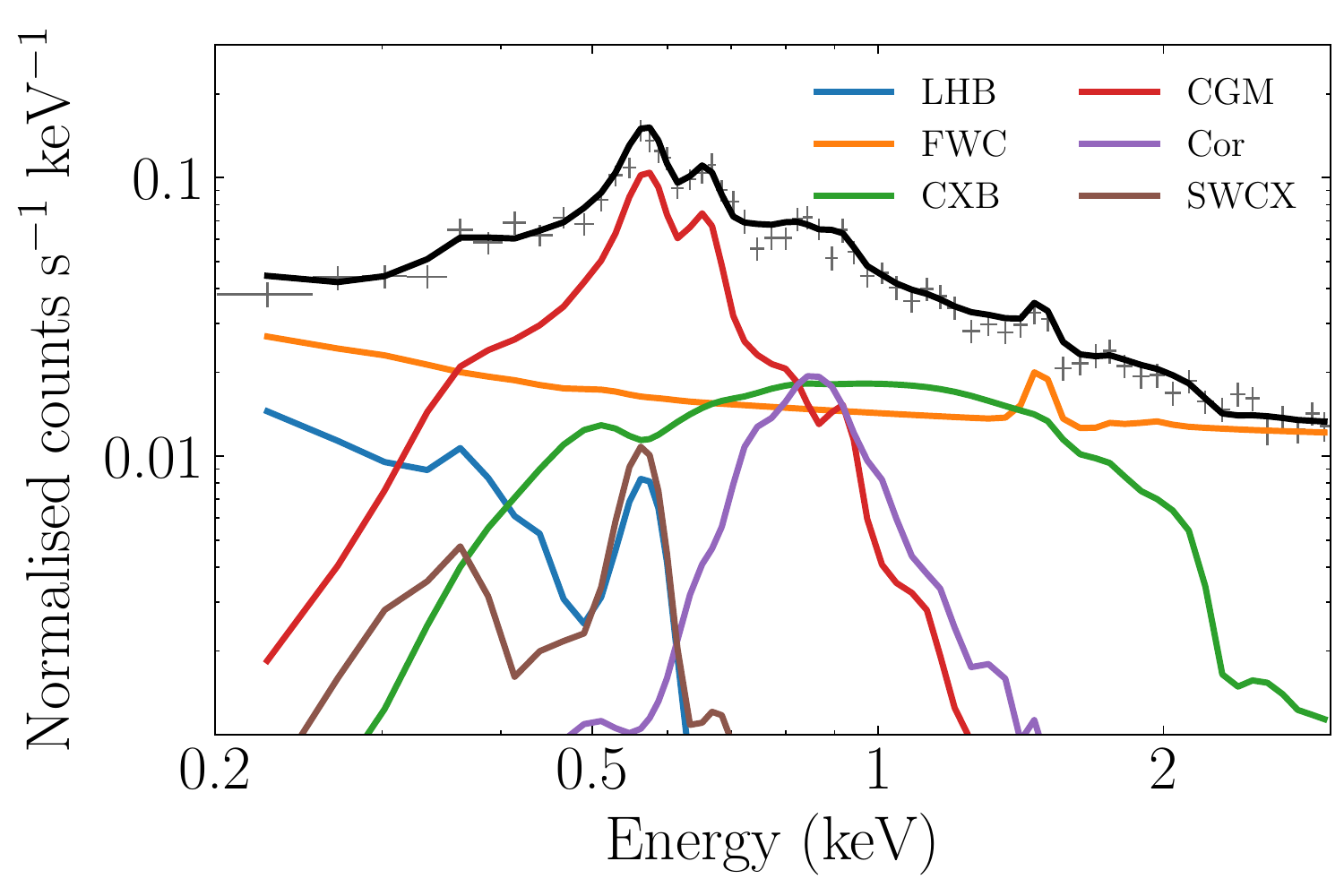}
    \caption{Illustration of all the spectral components taken from Cha~II \& III.}
    \label{fig:spec_demo}
\end{figure}

This section describes the various components that comprise the soft X-ray spectrum and our spectral fitting procedures. 
The spectral components are summarised in Fig.\,\ref{fig:spec_demo}, which we subsequently elaborate on individually in the rest of the section.

We divide the physical components into two groups, foreground and background. The former includes the local hot bubble (LHB) and solar wind charge exchange (SWCX) emissions in front of the molecular clouds. The latter comprises the circum-galactic medium (CGM), cosmic X-ray background (CXB) and a Galactic corona (Cor) component \citep{Ponti22}. The background components, for simplicity, are assumed to be absorbed by the same column within the cloud. In addition, we assume there is no absorption between us and the cloud.

The instrumental background is fixed using the empirical models developed from the filter-wheel closed data accumulated since the launch of SRG. Because the instrumental background differs slightly between the 5 TMs, as shown in Appendix\,\ref{sec:FWC}, the instrumental model used is TM-specific.
In Appendix\,\ref{sec:FWC}, we also demonstrate that the eROSITA instrumental background remained stable within the $0.2$--$9$\,keV, independent of time and CCD temperature.

\subsection{Foreground Components}
Chamaeleon\,II \& III are located at a distance of $\sim190$--$200$~pc \citep{voirin,Galli21} which lie on the surface of the LHB \citep{zucker}. In the simplified scenario, most, if not all, of the soft X-ray photons behind the cloud are absorbed, and one could isolate the foreground soft X-ray emissions due to the LHB and SWCX. As one would see later in Fig.\,\ref{fig:cha_fit} in Sect\,\ref{sec:result}, this is only realised in two regions with the highest column densities, below $\sim0.4$~keV, where the two foreground components begin to emerge over the CGM emission. The limited energy range where the foreground components dominate but remain below the instrumental background component necessitates using regions (1 and 2) of lower column densities to estimate the contribution of other background components below $\sim0.4$~keV.

We modelled the LHB as a plasma in collisional ionisation equilibrium \citep{Liu16}, which is described by the \texttt{APEC} model \citep{apec}. The LHB component is assumed to have solar abundance, given that the Sun is embedded within it. Therefore, the only free parameters of the LHB component are the plasma temperature and the emission measure (EM). We note that for each region, the normalisation of the LHB is scaled by the region area only and remains independent of time (eRASS). The latter reflects that the LHB does not vary on time scales from half a year to two years. As such, the LHB component is only scaled by one free EM parameter, despite different panels showing various combinations of region and eRASS as shown in Fig.\,\ref{fig:cha_fit}.
This single-parameter component normalisation applies to all other background components (CGM, corona and CXB), and the additional regions introduce no extra free parameter.

The other foreground component is SWCX, which would produce extra diffuse soft X-ray emission in the foreground. 
The primary source of SWCX in eROSITA is expected to be heliospheric. In heliospheric SWCX, the charge exchange between the neutral atoms in the ISM and ionised particles in the solar wind emits X-ray lines.
For magnetospheric SWCX, the solar wind is shocked, compressed, and interacts with the much denser neutral exosphere. The magnetospheric SWCX is expected to vary on a shorter time scale ($\sim$ minutes to days).
In eROSITA observations, the magnetospheric SWCX contribution is expected to be negligible (at least for high ecliptic latitude sources) since, being at L2, eROSITA observation geometry intersects only the most tenuous flanks of the magnetosheath. Another potential source of magnetospheric SWCX is the surface of the magnetotail. We believe this effect is minimal because, as shown in more detail in Appendix\,\ref{sec:orbit}, SRG/eROSITA was mostly located outside the magnetosphere due to its halo orbit around L2 during our observations, and additionally, no excess emissions have been found from the frequent observations through the magnetotail of the south ecliptic pole.


The time scale of the heliospheric SWCX is generally longer than the magnetospheric SWCX because the SWCX flux received is integrated along the line of sight up to the edge of the heliosphere, averaging out shorter variations of the solar wind. This is likely true for Cha~II \& III and CrA that are located at higher ecliptic latitudes ($\beta \sim -62\degr$ and $-14\degr$) and are thus out of the plane of the Parker spiral \citep{Parker}. However, the situation is less clear for Oph ($\beta \sim-1\degr$) near the ecliptic plane. A shorter variation time scale in order of hours to days is possible if the sight line is parallel to the pattern of the Parker spiral nearby and vice versa for the perpendicular case (Dennerl et al. in prep).

Ignoring the potential complication for Oph, a simple estimation using the distance to the heliopause ($\sim 120$~AU) and the mean solar wind speed of $ 450$~km\,s$^{-1}$ yields a time scale of $\sim 460$~days \citep{Kuntz2019}. However, the line-of-sight integration is heavily weighted by the $r^{-2}$-dependence of the solar-wind density away from the Sun, which for reference, is $\sim 1\%$ of its initial density at $r=10$\,AU --- a distance solar wind only takes $\sim 40$~days to traverse.  Therefore, the heliospheric SWCX likely vary from days to weeks. Recently, \citet{Qu2022} showed that the heliospheric SWCX is also correlated positively with the solar cycle using \ion{O}{VII} and \ion{O}{VIII} line fluxes measured as by XMM-Newton for $10$~years until 2010. Their study revealed a long-term variation in the heliospheric SWCX. However, it is not sensitive to shorter-term variations in the order of or less than half a year, as the width of the line flux bin was chosen to be half a year. SWCX variations on time scales of half a year between CalPV to eRASS3 have also been found by \citet{Ponti22}. In addition, within the $\sim4$~days scanning time of the eFEDS field, no noticeable variation from SWCX was found \citep{Patchiness}. The two studies support the expectation that eROSITA is only subject to the  heliospheric SWCX observing from L2.

In our spectral fitting, we modelled the SWCX component using the \texttt{ACX2} (v1.0.3) model \citep{ACX,ACX2}. The \texttt{ACX2} model supersedes the older version \texttt{ACX} by including velocity-dependent effects and charge exchange cross sections from the Kronos database \citep{Mullen2016,Mullen2017,Cumbee2018}. We simplify the model by assuming all the solar wind ions have a single velocity at the mean solar wind speed ($450$~km\,s$^{-1}$), the fraction of neutral Helium at the cosmic value 0.09 (default), solar abundance, recombination type to be single recombination and the \texttt{acxmodel} parameter to be 4. We found the choice of solar wind speed and the \texttt{acxmodel} parameter is not critical to the shape of the SWCX component by inspecting model \texttt{ACX2} spectra with eROSITA's spectral resolution, while for the other fixed parameters, we argue that they hold representative values. 

The SWCX component is allowed to vary from cloud to cloud primarily because (1) the clouds were observed at different times during eRASSs, and (2) SWCX is expected to be spatially variable on large angular scales. The normalisation of the SWCX component is allowed to vary for each eRASS to account for variability. However, for different regions within the same eRASS, the SWCX normalisation is fixed by the corresponding region area without introducing extra degrees of freedom. Last but not least, we simplify the model on the freeze-in temperature $T_{\mathrm{SWCX}}$. The freeze-in temperature sets the solar wind ion population by assuming the ion population was in collisional ionisation equilibrium with electrons at this temperature. We assume the freeze-in temperature would not change between eRASSs, and therefore, $T_{\mathrm{SWCX}}$ constitutes a single free parameter in the spectra fitting.


\subsection{Background Components}
We model the non-instrumental X-ray background with three model components: CXB, CGM, and Galactic corona. These components are all modulated by the absorption of the molecular cloud, which we model using the \texttt{tbabs} model \citep{tbabs}. It is important to note that we leave $N_\mathrm{H}$ as a free parameter and do not impose priors on $N_\mathrm{H}$ in all extraction regions, given the systematic uncertainties from various tracers as shown in Table\,\ref{tab:exposure}. The \textit{Herschel} $N_\mathrm{H}$ information only defines the regions. For simplicity, we take the approximation that all absorptions occur within the cloud, hence the same column density for all background components within the same region.

The presence of the CXB was first observed by \citet{Giacconi1962}. The CXB was subsequently found to be isotropic \cite[e.g][]{Schwartz}, suggesting an extragalactic origin. Nowadays, it is known that a range of sources, including active galactic nuclei, galaxies, and galaxy clusters, all contribute to the CXB \citep[see][for a review]{Brandt2021}. \citet{Cappelluti17} found that 91\% of the observed CXB can be resolved to detected X-ray sources and galaxies from the Chandra COSMOS-legacy field. The CXB spectrum can be modelled as a power law with $\Gamma$ $\sim 1.4$--$1.5$ above $\sim 1$~keV \citep{Vecchi,Kushino,Hickox,Cappelluti17}. \citet{Ponti22} also tested a double power law with $\Gamma_1=1.9$ below $0.4$~keV, $\Gamma_2=1.6$~keV, and $\Gamma_3=1.45$ above $1.2$~keV for the CXB component in light of the observational constraints on the CXB \citep{Gilli}. While the latter is likely more realistic, we justify using a simple power law with a fixed $\Gamma=1.45$ as most of the CXB is absorbed by the column density of the molecular cloud or subdominant to the CGM component. This value of $\Gamma$ is based on the measurement of \citet{Cappelluti17}, who found $\Gamma=1.45\pm0.02$. Upon fixing the photon index, the only parameter allowed to vary is the normalisation of the CXB component.

We model the emissions from the Milky Way as a combination of the CGM and Galactic corona, following the treatment of \citet{Ponti22}. The CGM is generally attributed to the hot gas halo of the Milky Way \citep[e.g.][and references therein]{Miller16,Ponti22}, which could extend up to its virial radius of $\sim  280$\,kpc or beyond \citep{MW}. This hot gas halo could be the shock-heated gas created by accretion onto Milky Way's dark matter halo \citep[e.g.][]{White91}. %
In our model, the CGM is assumed to be in collisional ionisation equilibrium like the LHB, but with a much lower abundance of 0.1~$Z_\odot$ compared to the conventional value of $Z_{\mathrm{CGM}}=0.3\,Z_\odot$. The low abundance is motivated by the finding of \citet{Ponti22} in the eFEDS field, who found $Z\simeq0.06\,Z_\odot$ with an upper limit of $Z\sim0.1\,Z_\odot$. 

The Galactic corona component was detected in the eFEDS field as well and is hypothesised to originate from energetic activities such as supernova explosions with sufficient energies to break free from the Galactic disk and supply hot plasma and metals into the spaces above and below the disk \citep[e.g.][]{Fraternali15, Ponti22}. If one assumes the corona is in collisional ionisation equilibrium with solar abundance, \citet{Ponti22} inferred a temperature of $\simeq 0.7$~keV. Solar abundance is assumed because the source of the corona is believed to be from chimneys or outflows from the Galactic disk, which are expected to be chemically enriched.

There is a concern that the corona component could be confused with the coronal emission from M dwarfs, as suggested by \citet{Wulf_2019}. A recent eROSITA study by \citet{Magaudda22} suggests 65\% of the 687 detected M dwarfs have a temperature $\sim 0.5$~keV, similar to the corona component. It is unclear at the moment the significance M dwarfs play in the soft X-ray diffuse emission, whether they are negligible, partly or wholly responsible for the corona component. However, analysing the physical origin of this component is beyond the scope of our work.

With the aforementioned assumptions, only temperatures and their EMs are free to vary in the CGM and corona components. The solar abundances reference for all components modelled by \texttt{APEC} (LHB, CGM, Cor) follows \citet{Anders}.

\subsection{Fitting Procedures}
Spectral fitting was performed with the PyXspec software \citep{xspec,pyxspec}. For each cloud, we carried out a simultaneous fit of 80 spectra (5 TMs $\times$ 4 regions $\times$ 4 eRASSs). The spectral fits began with minimising C-statistic \citep{Cash1979}. The resulting covariant matrix would then be used to construct a gaussian proposal distribution in the following Markov Chain Monte Carlo (MCMC) step. We ran three MCMC chains for each cloud using the Goodman-Weare method \citep{GW}, each with 100 walkers of 10000 steps. The initial positions of the walkers were randomised to avoid being trapped in local minima, and the first 1000 steps they sampled were discarded. As all the resulting corner plots (marginalised posterior distributions) from merging the three chains exhibit only a single peak without complicated profiles (see Appendix~\ref{appendix:post}), we finally constructed the model spectra using the 50 percentile of each parameter (as in Figs.\,\ref{fig:cha_fit}, \ref{fig:rcra_fit} and \ref{fig:oph_fit}). 

\section{Results and Discussions} \label{sec:result}

\begin{figure*}
    \centering
    \includegraphics[width=\textwidth]{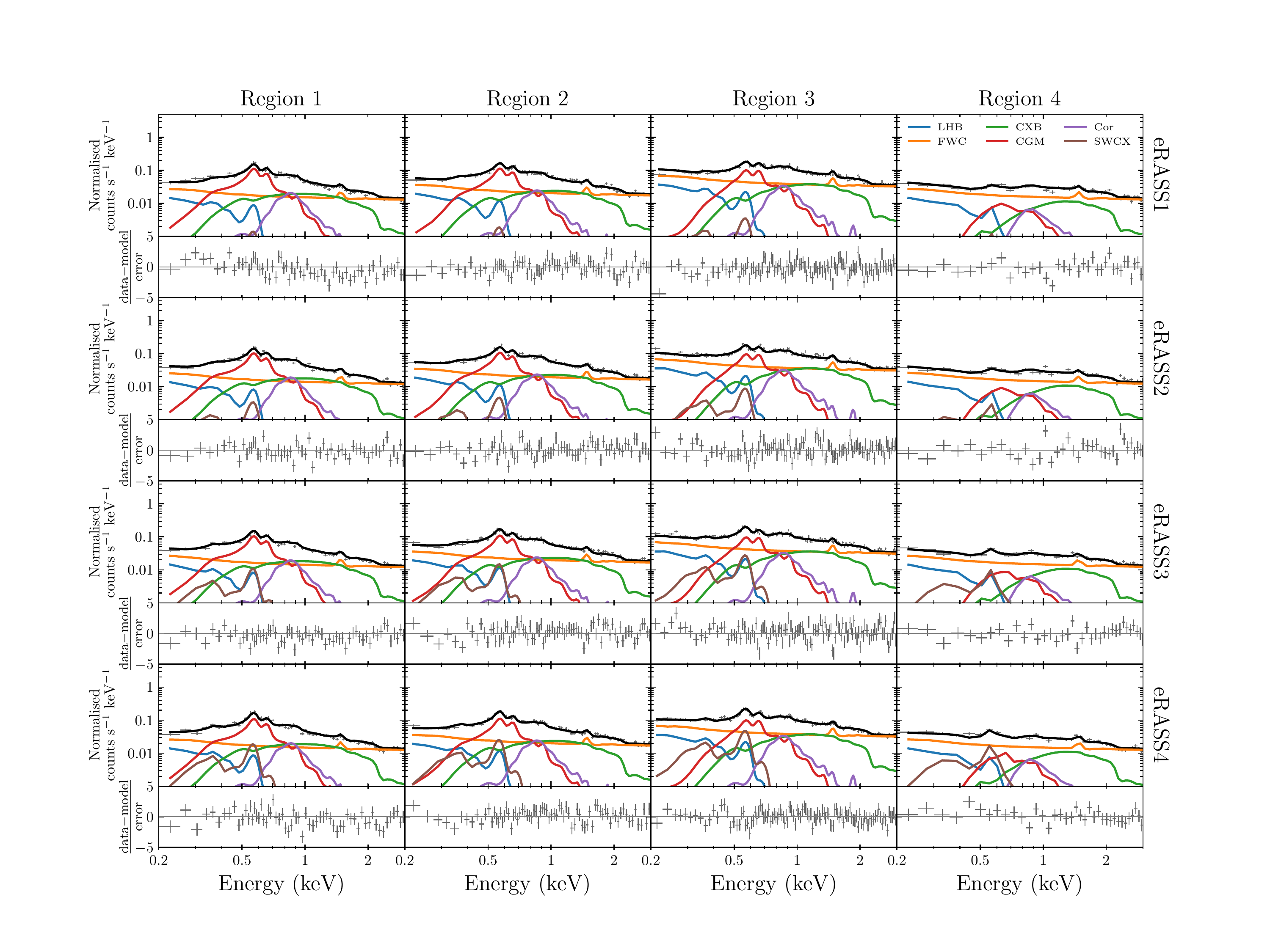}
    \caption{Spectral model of Cha~II \& III as functions of time in half-year intervals (from top to bottom row) and column density (from left to right column). Each panel shows the TM-averaged data in grey and the corresponding model in black. The constituents of the models are also shown (see the legend).}
     \label{fig:cha_fit}
\end{figure*}

\begin{figure*}
    \centering
    \includegraphics[width=\textwidth]{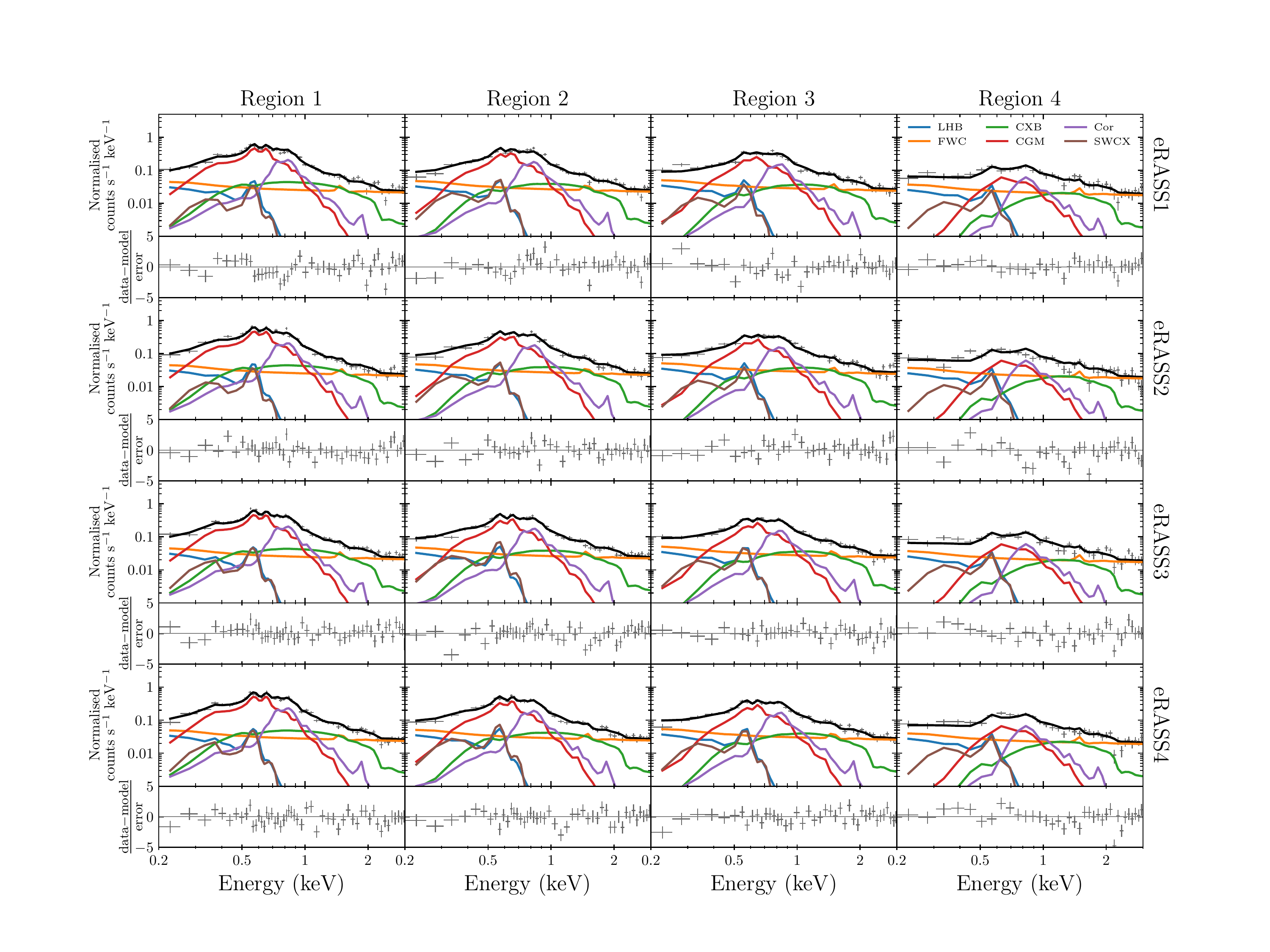}
    \caption{Same as Fig.\,\ref{fig:cha_fit}, but for CrA.}
     \label{fig:rcra_fit}
\end{figure*}

\begin{figure*}
    \centering
    \includegraphics[width=\textwidth]{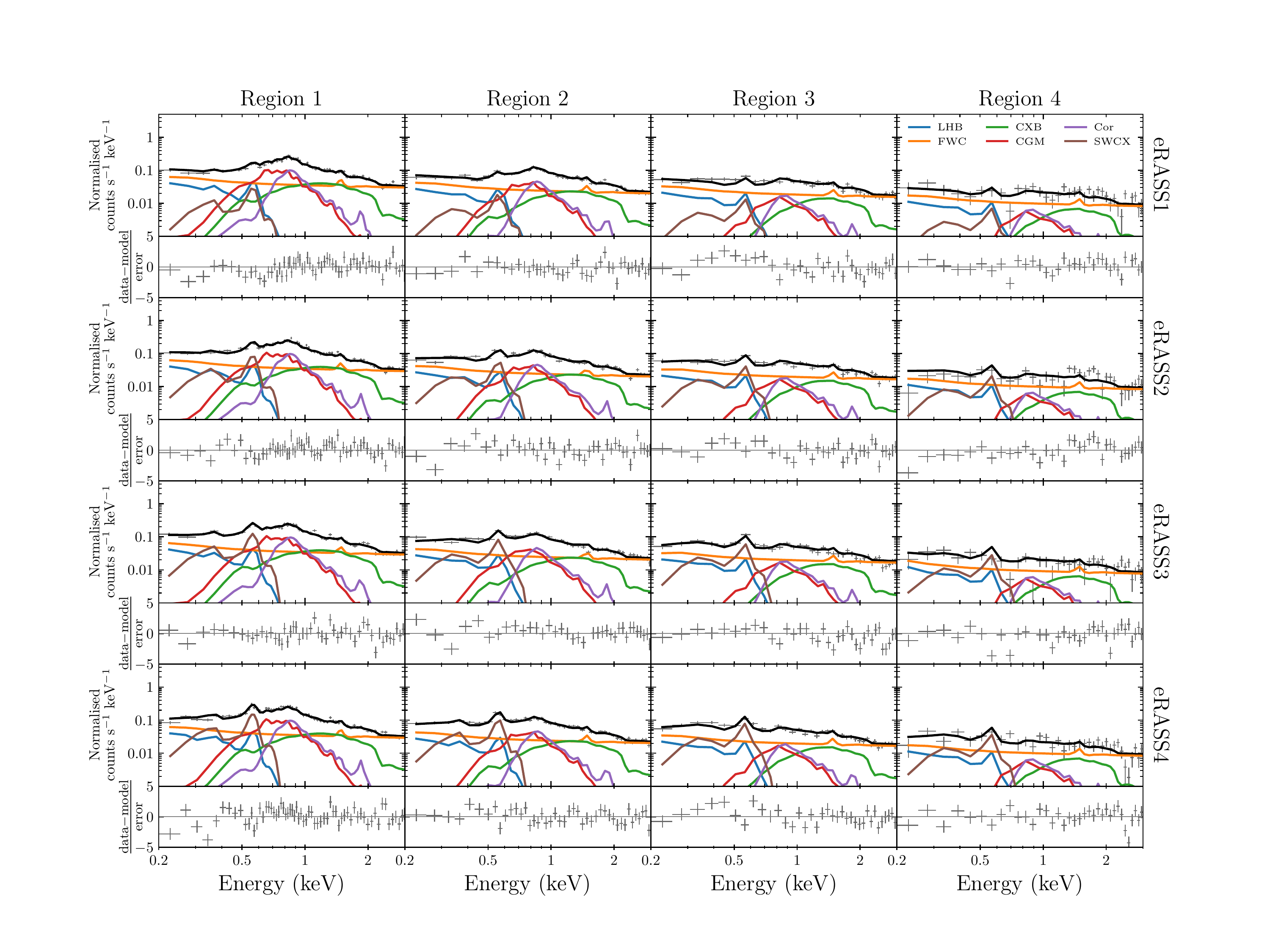}
    \caption{Same as Figs.\,\ref{fig:cha_fit} and \ref{fig:rcra_fit}, but for Oph.}
     \label{fig:oph_fit}
\end{figure*}

The fit with the highest likelihood for each cloud is shown in Figs.\,\ref{fig:cha_fit}, \ref{fig:rcra_fit} and \ref{fig:oph_fit} with the corresponding parameters in Table\,\ref{table:best-fit}. Figs.\,\ref{fig:cha_fit},  \ref{fig:rcra_fit} and \ref{fig:oph_fit} are arranged such that each row corresponds to the observation within the one eRASS, as indicated on the right axis. Each column corresponds to spectra extracted within the same region, as defined by the $N_\mathrm{H}$ maps as shown in the right panel of Figs.\,\ref{fig:img}, \ref{fig:rcra_img} and \ref{fig:oph_img}. Column density increases from left to right. We present the spectral fit in this manner so that the difference between the rows reflects the variations from  SWCX, and the difference between the columns shows the absorption of the background components. There are no strong correlations between the parameters, as evidenced by the two-dimensional projections of the posterior distributions shown in Appendix\,\ref{appendix:post}. 

\begin{table}
\caption{Fit parameters of the spectral fitting.}
\centering
\begin{tabular}{cccc}
\hline\hline
Cloud & Cha\,II \& III & Oph & CrA\\\hline
$kT_{\mathrm{LHB}}$\tablefootmark{(a)} & $0.084^{+0.004}_{-0.004}$ & $0.102^{+0.006}_{-0.006}$ & $0.115^{+0.012}_{-0.011}$ \\ 
$\mathrm{EM}_{\mathrm{LHB}}$\tablefootmark{(b)} & $2.563^{+0.356}_{-0.299}$ & $2.062^{+0.237}_{-0.221}$ & $1.913^{+0.369}_{-0.284}$ \\ 
$kT_{\mathrm{SWCX}}$\tablefootmark{(a)} & $0.109^{+0.003}_{-0.004}$ & $0.110^{+0.002}_{-0.002}$ & $0.108^{+0.005}_{-0.006}$ \\ 
$n_{\mathrm{SWCX, e1}}$\tablefootmark{(c)} & $0.192^{+0.217}_{-0.159}$ & $1.687^{+0.511}_{-0.530}$ & $3.215^{+1.297}_{-1.182}$ \\ 
$n_{\mathrm{SWCX, e2}}$\tablefootmark{(c)} & $0.495^{+0.306}_{-0.230}$ & $4.948^{+0.616}_{-0.617}$ & $3.271^{+1.207}_{-1.177}$ \\ 
$n_{\mathrm{SWCX, e3}}$\tablefootmark{(c)} & $1.544^{+0.428}_{-0.305}$ & $7.214^{+0.783}_{-0.811}$ & $4.200^{+1.235}_{-1.102}$ \\ 
$n_{\mathrm{SWCX, e4}}$\tablefootmark{(c)} & $2.596^{+0.540}_{-0.426}$ & $8.635^{+0.915}_{-0.834}$ & $3.973^{+1.110}_{-1.232}$ \\ 
$N_{\mathrm{H,reg1}}$\tablefootmark{(d)} & $1.106^{+0.062}_{-0.059}$ & $2.618^{+0.191}_{-0.162}$ & $0.641^{+0.050}_{-0.044}$ \\ 
$N_{\mathrm{H,reg2}}$\tablefootmark{(d)} & $1.434^{+0.066}_{-0.054}$ & $3.665^{+0.191}_{-0.177}$ & $1.196^{+0.063}_{-0.052}$ \\ 
$N_{\mathrm{H,reg3}}$\tablefootmark{(d)} & $2.313^{+0.070}_{-0.058}$ & $5.566^{+0.215}_{-0.220}$ & $1.673^{+0.067}_{-0.062}$ \\ 
$N_{\mathrm{H,reg4}}$\tablefootmark{(d)} & $4.258^{+0.124}_{-0.085}$ & $6.582^{+0.415}_{-0.370}$ & $3.324^{+0.118}_{-0.113}$ \\ 
$kT_{\mathrm{CGM}}$\tablefootmark{(a)} & $0.183^{+0.002}_{-0.002}$ & $0.270^{+0.010}_{-0.011}$ & $0.213^{+0.005}_{-0.005}$ \\ 
$\mathrm{EM}_{\mathrm{CGM}}$\tablefootmark{(e)} & $6.950^{+0.437}_{-0.478}$ & $5.581^{+0.851}_{-0.745}$ & $12.529^{+0.799}_{-0.673}$ \\ 
$kT_{\mathrm{Cor}}$\tablefootmark{(a)} & $0.746^{+0.017}_{-0.014}$ & $0.719^{+0.017}_{-0.018}$ & $0.613^{+0.014}_{-0.015}$ \\ 
$\mathrm{EM}_{\mathrm{Cor}}$\tablefootmark{(b)} & $0.670^{+0.034}_{-0.039}$ & $2.294^{+0.222}_{-0.226}$ & $3.272^{+0.211}_{-0.226}$ \\ 
$\mathrm{norm}_{\mathrm{CXB}}$\tablefootmark{(f)} & $0.237^{+0.003}_{-0.003}$ & $0.286^{+0.006}_{-0.007}$ & $0.278^{+0.008}_{-0.008}$ \\ 
\hline
\end{tabular}

\label{table:best-fit}
\tablefoot{The values reported are the $50$ percentiles, with the lower and upper bounds showing the $16$ and $84$ percentiles of the Markov Chain Monte Carlo analysis result.\\
\tablefoottext{a}{$kT_{\mathrm{LHB}}$, $kT_{\mathrm{SWCX}}$, $kT_{\mathrm{CGM}}$ and $kT_{\mathrm{Cor}}$ are in units of keV.} \\
\tablefoottext{b}{$\mathrm{EM}_{\mathrm{LHB}}$ and $\mathrm{EM}_{\mathrm{Cor}}$ are in units of $10^{-3}\,{\rm cm^{-6}\,pc}$.}\\
\tablefoottext{c}{$n_{\mathrm{SWCX}}$ is in the unit of $10^{-2}\,\mathrm{deg}^{-2}$. The normalisation parameter of the \texttt{ACX2} model is dimensionless and is only intended for relative scaling (see the documentation of the \texttt{ACX} model). We normalised this factor by the sky area to give the unit deg$^{-2}$.}\\
\tablefoottext{d}{$N_{\mathrm{H}}$ values are in units of $10^{21}\,\mathrm{cm^{-2}}$.}\\
\tablefoottext{e}{$\mathrm{EM}_{\mathrm{CGM}}$ is in $10^{-2}\,{\rm cm^{-6}\,pc}$.}\\
\tablefoottext{f}{$\mathrm{norm}_{\mathrm{CXB}}$ has unit of $10^{-2}$\,photons\,keV$^{-1}$\,cm$^{-2}$\,s$^{-1}$\,deg$^{-2}$ at 1 keV.} 
}
\end{table}

\subsection{SWCX} \label{subsec:SWCX}
For Cha~II \& III, one can immediately identify the monotonically increasing SWCX component from eRASS1 to eRASS4, in line with the observations in eFEDS and eRASSs (\citealt{Ponti22}, Dennerl et al. in prep.). The SWCX contribution is negligible in eRASS1, matches the \ion{O}{VII} line from the LHB subsequently at eRASS3, then ultimately dominates the foreground \ion{O}{VII} emission in eRASS4. This trend echoes the natural expectation and the findings by \citet{Qu2022}, where the heliospheric SWCX is positively correlated with the solar cycle (solar cycle 25 began on December 2019, as well as eRASS1). 

We found the best-fit $kT_\mathrm{SWCX}\sim 0.1$\,keV for all three sight lines, a value consistent with solar wind data \citep{von_Steiger_2000,Gloeckler2007}. Despite similar temperatures of the LHB and SWCX ($kT_{\mathrm{LHB}}\simeq kT_{\mathrm{SWCX}}\simeq0.1$~keV), the spectral shapes of the SWCX and LHB are significantly different --- SWCX has two prominent peaks, \ion{C}{VI} at $0.37$~keV and \ion{O}{VII}  at $0.57$~keV, while the LHB has both of these prominent emission lines, an extra rising continuum towards the low energy from radiative recombination and bremsstrahlung is present. Combining such difference with the fact that only the SWCX component is allowed to vary between eRASSs, eliminates much of the degeneracy between the two components, given our SWCX spectral model is correct and our constant $kT_{\mathrm{SWCX}}$ assumption is valid. The (lack of) correlations between the LHB and SWCX parameters can be shown from the posterior distributions of the model parameters in Fig\,\ref{fig:corner}.

\begin{figure}[htbp]
    \centering
    \includegraphics[width=0.45\textwidth]{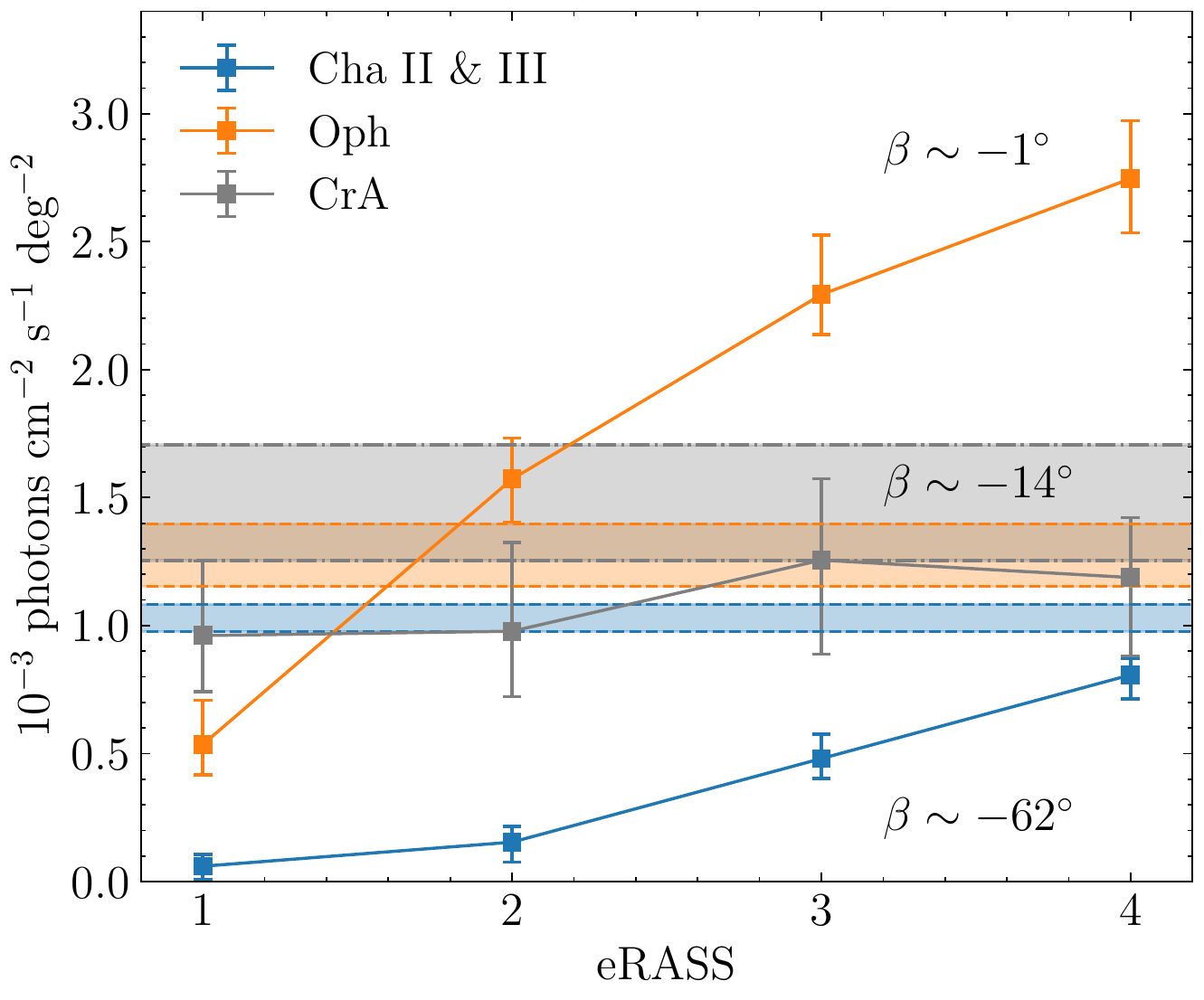}
    \caption{Variation of the model SWCX intensity in the 0.3--0.7~keV band in eRASS1--4. The LHB intensities within $1\,\sigma$ confidence level in the same band are also shown by the shaded regions.}
    \label{fig:swcx}
\end{figure}

Fig.\,\ref{fig:swcx} summarises the SWCX variability to the directions of the three clouds. It shows the 0.3--0.7~keV band intensities from the models as a function of eRASS. This energy band includes most of the emissions from our SWCX component. Clear increasing trends are observed for all the clouds, showing the correlation with the solar activity irrespective of the pointing direction. Interestingly, among the three clouds, we observe that the SWCX intensity is the highest towards Oph and the lowest towards Cha~II \& III. We argue this is likely the difference in the zero-level SWCX, caused by the difference in the solar wind density, which is a decreasing function of heliolatitude \citep[e.g. see Fig. 18 of][]{Porowski22}. Ignoring the minor difference between heliolatitude and ecliptic latitude ($\beta$)\footnote{The solar equatorial plane is inclined by $\sim 7\fdg5$ with the ecliptic plane, with the line of nodes at ecliptic longitude of $\sim 76\degr$ \citep{almanac,Hapgood1992}.}, one could see this argument matches, at least qualitatively, with the ecliptic latitude of the clouds --- Cha~II \& III has the highest $\beta\sim-62\degr$, followed by CrA at $\beta\sim-14\degr$, and finally Oph at $\beta\sim-1\degr$. As a result, the LHB intensity dominates the SWCX emissions from Cha~II \& III in all eRASSs in this band, but SWCX overtook it towards Oph since eRASS2.

We notice that the rate of increase of SWCX intensity is not at all constant and differs from cloud to cloud. Explaining the precise trends is complex and is beyond the scope of this work. To complicate the issue further, we point out that Fig.\,\ref{fig:swcx} contains a caveat where the observations of different clouds made within the same eRASS were not done simultaneously. Instead, they could differ up to $\sim 1.5$ months in the case of CrA ($\lambda \sim 282\degr$) and Cha~II \& III ($\lambda \sim 246\degr$), because eROSITA's scanning approximately follows the ecliptic longitude ($\lambda$) with a progression of about 1\degr per day. 

The background $0.3$--$0.7$\,keV count rate of RX~J1856 located in the neighbourhood of CrA indicates a similar trend as CrA from eRASS1--3 (see Table\,\ref{tab:rxj_rate}). This comparison suggests our inferred SWCX contribution from spectral fitting is reliable, and indeed the increase in SWCX emission from CrA is less pronounced than the rest (see Appendix\,\ref{appendix:rxj} for more discussion).

\subsection{LHB} \label{subsec:LHB}
We found the temperature of the LHB to range from $0.084^{+0.004}_{-0.004}$--$0.115^{+0.012}_{-0.011}$~keV from the three sight lines. We also found EM$_{\mathrm{LHB}}$ to span a range from $1.91^{+0.37}_{-0.28}$ -- $2.56^{+0.36}_{-0.30}\times10^{-3}$~cm$^{-6}$\,pc. As mentioned in \citet{Bluem2022}, a direct comparison of these values, which are derived from the \texttt{APEC} model in \texttt{AtomDB} version 3.0.9, with those measured by \citet{Liu16} is no longer valid due to the updates introduced since its publication date. Therefore, we repeat the same spectral fitting using \texttt{AtomDB} version 3.0.3, released about half a year before the publication of \citet{Liu16}. This result is shown in more detail in Appendix\,\ref{sec:apec_v303}. In summary, the LHB temperature we found is consistent with the SWCX-subtracted measurement from ROSAT by \citet{Liu16}, who found $kT_{\mathrm{LHB}}=0.097\pm0.019$~keV averaging across the whole sky. In terms of EM, our measurement is in line with values inferred from the EM map of \citet{Liu16}.
One could also compare our LHB measurement to \citet{McCammon2002}, who observed the diffuse X-ray background using microcalorimeters with a high energy resolution of $5$--$12$\,eV on a sounding rocket. \citet{McCammon2002} reported a $kT_\mathrm{LHB}$ of $0.099$\,keV and an $\mathrm{EM}_{\mathrm{LHB}}$ of $8.8\times 10^{-3}$\,cm$^{-6}$\,pc --- a similar temperature but a significantly different $\mathrm{EM}_{\mathrm{LHB}}$. The main sources of the discrepancy are likely to be a higher Galactic latitude in their observations ($|\Delta b| \sim 40\degr$), a limited knowledge of SWCX at the time, which was not modelled, as well as a significantly fainter Galactic halo component ($\mathrm{EM}_{\mathrm{CGM}} = 3.7\times10^{-3}\,$cm$^{-6}$\,pc) in their spectral fits.

Using the Wisconsin B/C band intensity ratio, \citet{Snowden1990} reported a dominantly longitudinal LHB temperature dipole, from $10^{5.9}$\,K towards the Galactic anti-centre to $10^{6.2}$\,K towards the Galactic centre. Later, the existence of the temperature dipole was further consolidated using X-ray shadows and their ROSAT R2/R1 band ratio by \citet{Snowden2000}, which show a minimum temperature of $10^{6.04}$~K ($0.09$~keV) near the Galactic anti-centre and a maximum of $10^{6.13}$~K ($0.11$~keV) at the Galactic centre. We observe the same trend and systematic temperature variation in the three molecular clouds, which shows a decreasing temperature from $0.084^{+0.004}_{-0.004}$\,keV at $l\sim300\degr$ to $0.115^{+0.012}_{-0.011}$~keV at $l \sim 360\degr$, although the X-ray shadows in \citet{Snowden2000} are located at higher latitudes ($|b| \gtrsim 20\degr$). We would like to note it is not clear whether this is a real temperature change, as incorrect modelling of the absorption could also result in an apparent temperature change in the LHB.

It is also interesting to test if the LHB is homogeneous in density with known distances of the clouds in literature. The distances to these clouds were mainly derived from \emph{Gaia}-DR2 astrometry of young stars embedded within the cloud complexes. For instance, the distances to Cha\,II \& III are known to a very high accuracy from \emph{Gaia} parallax measurements --- Cha~II is $197.5^{+1.0}_{-0.9}$~pc away \citep{Galli21} and Cha~III $199^{+20}_{-18}$~pc \citep{voirin}. 
{One could infer the electron density $n_e$ of the LHB using
\begin{eqnarray}
    \mathrm{EM} &=& \int^{L}_{0} n_e(l) n_{\element{H}}(l) dl,
\end{eqnarray}
 where $n_{\element{H}}$ is the hydrogen density and $L$ is the distance to the molecular cloud concerned.
We simplify the expression further by adopting a few assumptions --- the LHB is completely volume-filling up to the clouds with a constant density and it is fully ionised with $n_{\element{He}}$/$n_{\element{H}} = 0.1$, so that the electron density is $1.2$ times the hydrogen density $n_{\element{H}}$ of the LHB plasma \citep{Snowden14}. Therefore, we have
\begin{eqnarray}
   n_e = \sqrt{\frac{1.2\mathrm{EM}}{L}}.
\end{eqnarray}
We assume further that the distance to both cloud complexes is $L \simeq 198$~pc with negligible uncertainty compared to the EM, we infer $n_e = 3.94^{+0.26}_{-0.24}\times10^{-3}~\mathrm{cm}^{-3}$.} This dwells on the low side, but is nonetheless consistent with the measurement of $n_e=(4.68\pm0.47)\times10^{-3}$\,cm$^{-3}$ by \citet{Snowden14}.
Similarly, we estimate the electron density of CrA and Oph using the calculation above using their most recent distance measurements and present the results in Table\,\ref{tab:ne}. 


\begin{table*}
    \centering
    \begin{tabular}{cccccc}
    \hline \hline
        Cloud & Distance (pc) & $n_e$ ($10^{-3}$~cm$^{-3}$) & $P_T/k$ ($\mathrm{cm^{-3}\,K}$) & Distance references\\\hline
        Cha~II \& III & $197.5^{+1.0}_{-0.9}$ \& $199^{+20}_{-18}$ & $3.94^{+0.26}_{-0.24}$ & $7380^{+860}_{-780}$ & 1, 2 \\
        Oph & $141.2^{+8.4}_{-7.5}$ & $4.19^{+0.23}_{-0.23}$ & $9520^{+1110}_{-1050}$& 3\\
        CrA & $149.4^{+0.4}_{-0.4}$ & $3.92^{+0.36}_{-0.30}$ & $10050^{+2070}_{-1660}$ & 4\\\hline
    \end{tabular}
    \caption{Distances to the clouds and the estimation of the LHB electron density and thermal pressure along these sight lines.}
    \label{tab:ne}
    \tablebib{(1)~\citet{Galli21}; (2) \citet{voirin}; (3) \citet{Canovas}; (4) \citet{Galli20}}
    \tablefoot{The uncertainties in the distance are relatively small compared to the EM uncertainties reported in Table\,\ref{table:best-fit}. Therefore, they are ignored in the conversion to $n_e$.}
\end{table*}

\begin{figure}
    \centering
    \includegraphics[width=0.45\textwidth]{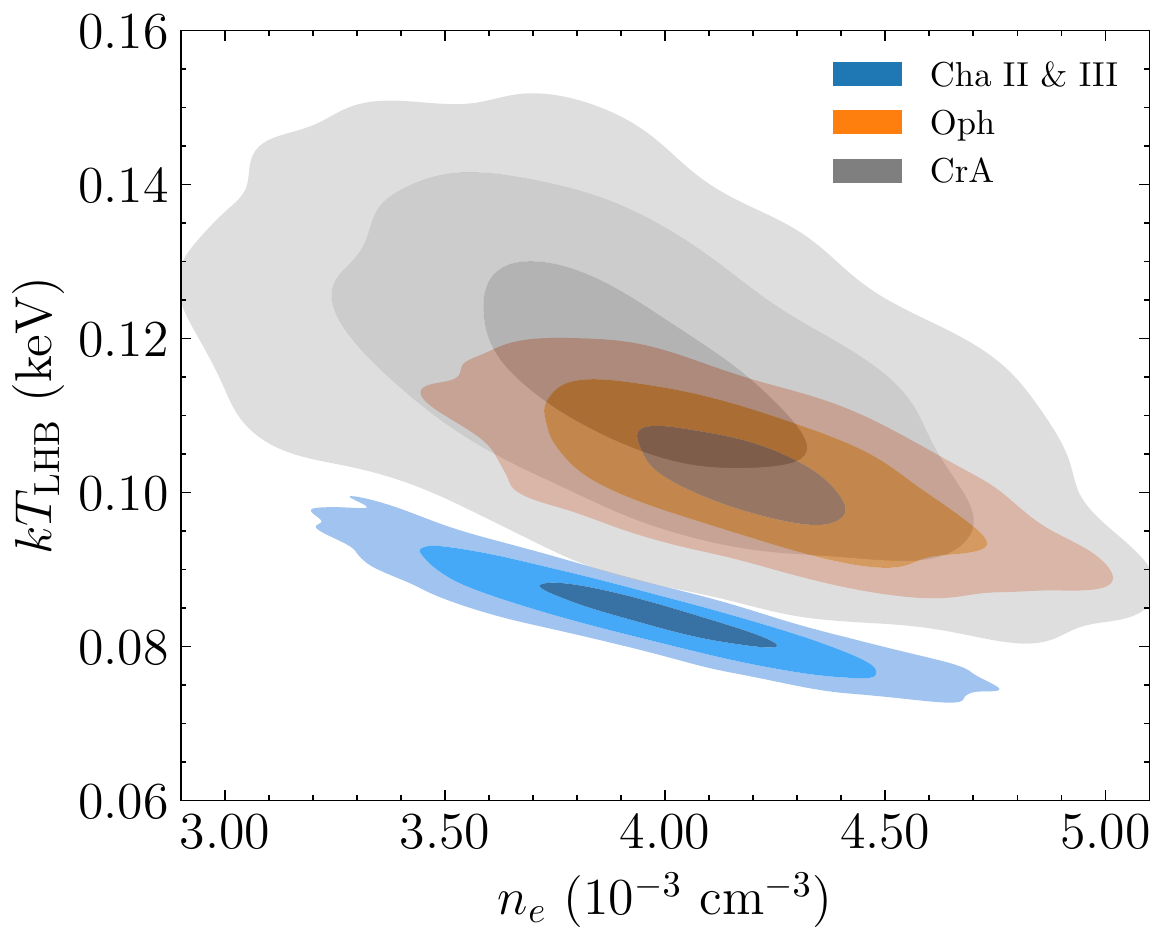}
    \caption{Posterior distributions of the temperature and electron density of the LHB. The contours indicate the 1, 2 and 3\,$\sigma$ confidence levels (enclose $\sim$\,39, 86 and 99\% of the probability from the highest density).}
    \label{fig:temp_vs_den}
\end{figure}

With the conversion from EM to electron density, we present the posterior distributions of the LHB properties in Fig.\,\ref{fig:temp_vs_den}. Despite the three lines of sight agreement on the LHB density at $\sim 4\times10^{-3}$~cm$^{-3}$, the LHB temperature inferred from Cha~II \& III is significantly colder than the rest. The LHB properties towards Oph and CrA agree at $1\,\sigma$ level but differ from those derived from Cha~II \& III at $\geq 3\,\sigma$ level. The variation in electron density of the three lines of sight is a constant within $10$\%.

We would like to highlight that the three chosen sight lines of significantly different distances (Cha~II \& III is $\sim 40\%$ farther than the rest) corroborate a single LHB density within $10$\% is an interesting result. This result strongly indicates that the LHB extends up to the molecular clouds' distance unless the LHB density or the volume filling factor is much more variable than we assumed.

The presence of a temperature gradient is not completely surprising as magneto-hydrodynamic (MHD) simulations have already shown the turbulent temperature and density structures sustained by supernova explosions in the ISM (e.g. see Fig.\,1 of \citealt{Hill14}, \citealt{Avillez2005}). It is reasonable to find varying temperature densities when integrating the emission in the turbulent LHB medium along different lines of sight. The fact that we inferred consistent LHB properties from Oph and CrA, which are separated by a significant angular separation of $\sim 35$\degr, indicates the constant temperature and density assumption is a reasonable modelling simplification to the more complicated scenario presented by MHD simulations.

We estimate the thermal pressure inside the LHB along the three lines of sight following the treatment in \citet{Snowden14} using
\begin{eqnarray}
   P_{T}/k = nT \simeq 1.92n_eT, 
\end{eqnarray}
 where $k$ is the Boltzmann constant and $n$ is the total particle density.
We found $P_T/k$ to be in the range of $6600$--$12100\,\mathrm{cm^{-3}\,K}$ (see Table\,\ref{tab:ne}). The resulting pressures are consistent with a constant, unlike the plasma temperature, as a result of propagating the uncertainties of $n_e$ and $T$. However, we note that there is a hint of pressure inhomogeneity between Cha~II \& III and the other two clouds ($\sim 1\,\sigma$), which could indicate a recent supernova or stellar wind expanding from within the LHB that has not yet attained pressure equilibrium with the ambient medium.

The LHB is more elongated in the direction perpendicular to than along the Galactic plane, most probably due to the higher pressure exerted by the ISM in the Galactic disk \citep{Liu16}. If the LHB is expanding adiabatically, we would expect a lower pressure also towards the Galactic poles. Unfortunately, our current samples cannot probe the Galactic-latitudinal dependence comprehensively as they are close to the Galactic plane and are only valid within $13\degr \lesssim |b| \lesssim 22\degr$.


\subsection{Background Components}
As we are probing the regions with some of the highest column density away from the Galactic plane, naturally, our analysis is not the best suited to the analysis of the background components located beyond $\sim200$~pc from us. Nonetheless, even from the images (left panels of Figs.\,\ref{fig:img}, \ref{fig:rcra_img} and \ref{fig:oph_img}), one could see the emissions from the eROSITA bubbles would bias our conclusion about the background components. Therefore, we begin our discussion with Cha~II \& III, the only cloud in our samples not located in front of the eROSITA bubbles, hopefully giving more reliable results.

It is instructive to begin by comparing the normalisation of the isotropic CXB component with literature values. With a single power-law with $\Gamma=1.48$, \citet{Chen1997} found $\mathrm{norm}_{\mathrm{CXB}}=10.5\pm0.4$~photons\,cm$^{-2}$\,s$^{-1}$\,sr$^{-1}$ at 1~keV, corresponding to $(3.2\pm0.2)\times10^{-3}$~photons\,cm$^{-2}$\,s$^{-1}$\,deg$^{-2}$, using a combination of ROSAT and ASCA. More recently, \citet{Cappelluti17} found  $\mathrm{norm}_{\mathrm{CXB}}=(3.32\pm0.05)\times10^{-3}$~photons\,cm$^{-2}$\,s$^{-1}$\,deg$^{-1}$ at 1~keV from the much more sensitive $2.15\,\mathrm{deg}^2$ Chandra COSMOS-legacy field data. Upon removing all the X-ray detected sources, they found $\mathrm{norm}_{\mathrm{CXB}}$ drops to $(1.27\pm0.08) \times10^{-3}$~photons\,cm$^{-2}$\,s$^{-1}$\,deg$^{-1}$ at 1~keV, corresponding to a decrease of $\sim2.5$ times. Our measurement of $\mathrm{norm}_{\mathrm{CXB}}$ for Cha~II \& III, but also the other two clouds, fall between the values of \citet{Cappelluti17} before and after the X-ray detected source removal. This is expected as eROSITA is much less sensitive and has a lower angular resolution than Chandra's 4.6 Ms COSMOS-Legacy survey, naturally detecting much less detected X-ray sources to exclude in our analysis.

We found the CGM component to be similar in temperature and EM as the value inferred from the eFEDS field\footnote{\citet{Ponti22} provides several models. Here, we refer to their shift-LHB-CGM-Coro-CXB-SWCX model, which is the most similar to our model set-up.} \citep{Ponti22}, despite the completely different line-of-sight. The eFEDS field and Cha~II \& III are separated by $\sim 85\degr$ in different galactic hemispheres and latitudes so that it could be a glimpse of an approximately spherical Milky Way hot halo preferred by the analysis of \citet{Bregman18}, but it is certainly far from conclusive.
Similarity can also be found in the temperature of the corona component ($\sim 0.7$~keV) when compared with the thermal equilibrium model in \citet{Ponti22}. However, we found a EM$_{\mathrm{Cor}}$ that is higher than the eFEDS field (EM$_{\mathrm{eFEDS}} =  (0.385\pm{0.025}) \times10^{-3}\,\mathrm{cm}^{-6}\,\mathrm{pc}$). This is qualitatively consistent with the picture that the corona is more confined to the Galactic plane from outflows from supernovae or stellar winds since Cha~II \& III has a lower galactic latitude than the eFEDS field ($|b|\sim 30\degr$).

A high galactic latitude ($|b| > 30\degr$) HaloSat study by \citet{Bluem2022} also detected the corona component, in which they call the hot component of the CGM, to have the same temperature $kT_{\mathrm{Cor}}\sim0.7$~keV. They inferred the same temperature for the cooler CGM component ($kT_{\mathrm{CGM}}\sim0.18$~keV) as our Cha~II \& III sight line. However, our assumptions of the abundances differ --- both their cool and hot CGM components are assumed to have $Z=0.3\,Z_\odot$. Regarding the EM, our values of both CGM and corona are significantly higher than those found by \citet{Bluem2022} in the closest sight lines (separated $\sim 15\degr$ from Cha~II \& III along the galactic longitude). The reason for the discrepancy is not apparent except for the different assumed abundance of both components. One possible explanation could be that the corona component is flattened along the Galactic plane, so the path length through a flattened corona increases more significantly towards the Galactic plane, resulting in a steep increase in EM.


For Oph and CrA, which are situated in front of the eROSITA bubbles, one could easily observe their spectra are much more enhanced around $\sim0.8$--$1$~keV, where the \ion{Ne}{IX}, \ion{Ne}{X} and various \element{Fe} lines are. The enhancement is most certainly coming from the eROSITA bubbles, which causes the EM of the corona component to $3$--$5$ times the value in Cha~II \& III. Naturally, the bubbles affect not only the fit to the corona component but also the CGM component, as their spectra largely overlap. The CGM is hotter towards Oph, while a much brighter but cooler CGM is preferred towards CrA. There are at least two possibilities that could contribute to the difference: (1) Oph has a higher column density in general, which could bias towards a higher CGM temperature because the temperature proxy \ion{O}{VIII}/\ion{O}{VII} line ratio could also be accommodated for by a higher column density; (2) the properties of the northern and southern eROSITA bubbles are different, where the northern bubble is likely hotter. The second possibility can be explored and constrained relatively tightly with large regions within the northern and southern eROSITA bubbles at much lower column densities, which would largely remove the ambiguity caused by significant opacity. Nevertheless, the eROSITA bubbles are not expected to affect the measured properties of the foreground components because of the high absorption.

\section{Summary and Conclusions} \label{sec:con}
We performed X-ray shadowing experiments on three giant molecular clouds using the data from the first four eRASSs. eROSITA spectra of the clouds allow us to separate the heliospheric SWCX and LHB contributions in soft X-ray foreground emission. We observed a monotonic increase of SWCX since eRASS1, independent of the cloud, matching the expectation based on the solar cycle. An ecliptic latitudinal dependence is also observed, consistent with the expected decreasing solar wind ion density. 

From the known distance to the clouds, we found a constant electron density of the LHB plasma towards all three clouds, with $n_e\sim4 \times10^{-3}$~cm$^{-3}$. However, we measured a lower LHB temperature towards Cha~II \& III ($kT_{\mathrm{LHB}}=0.084^{+0.004}_{-0.004}$~keV) compared to Oph and CrA ($\simeq 0.1$~keV). Our results show for the first time that there is a possible galactic longitudinal, but not a latitudinal gradient in the LHB temperature after the subtraction of SWCX. We also found that the thermal pressure in the LHB is consistent with constant given the current measurement uncertainty.

\begin{acknowledgements}
MY and MF acknowledge support from the Deutsche Forschungsgemeinschaft through the grant FR 1691/2-1.
GP acknowledges funding from the European Research Council (ERC) under the European Union’s Horizon 2020 research and innovation programme (grant agreement No 865637). MS acknowledges support from the Deutsche Forschungsgemeinschaft through the grants SA 2131/13-1, SA 2131/14-1, and SA 2131/15-1.

We would like to thank the anonymous referee for the constructive and in-depth comments. We thank Susanne Friedrich for sharing the results of the SEP orbital analysis prior to publication.

This work is based on data from eROSITA, the soft X-ray instrument aboard SRG, a joint Russian-German science mission supported by the Russian Space Agency (Roskosmos), in the interests of the Russian Academy of Sciences represented by its Space Research Institute (IKI), and the Deutsches Zentrum für Luft- und Raumfahrt (DLR). The SRG spacecraft was built by Lavochkin Association (NPOL) and its subcontractors, and is operated by NPOL with support from the Max Planck Institute for Extraterrestrial Physics (MPE).

The development and construction of the eROSITA X-ray instrument was led by MPE, with contributions from the Dr. Karl Remeis Observatory Bamberg \& ECAP (FAU Erlangen-Nuernberg), the University of Hamburg Observatory, the Leibniz Institute for Astrophysics Potsdam (AIP), and the Institute for Astronomy and Astrophysics of the University of Tübingen, with the support of DLR and the Max Planck Society. The Argelander Institute for Astronomy of the University of Bonn and the Ludwig Maximilians Universität Munich also participated in the science preparation for eROSITA.

The eROSITA data shown here were processed using the eSASS/NRTA software system developed by the German eROSITA consortium. 
\end{acknowledgements}

\bibliographystyle{aau}
\bibliography{ref}

\begin{appendix}

\section{Cross-checking the SWCX variation with XMM-Newton's  routine calibration source RX~J1856.5-3754} \label{appendix:rxj}
We would like to have an indicator of SWCX variability independent of our spectral analysis. Therefore, we used the EPIC-MOS2 data on board XMM-Newton of the routine calibration source RX~J1856.5-3754 (hereafter RX~J1856). RX~J1856 is a neutron star revisited by XMM-Newton approximately every six months for calibration. For the four observations carried out during the eRASSs, each has an exposure time ranging from $71500$--$74600$~s. These observations provide an independent constraint on the SWCX level because of its proximity to CrA, as indicated by the left panel of Fig.\,\ref{fig:rcra_img} with ($l$,$b$) = ($358\fdg59962$, $-17\fdg21311$). Both eROSITA and XMM-Newton require the solar angle to stay within $90\pm20$\degr, owing to operational considerations such as ensuring a sufficient power supply from the solar panels and thermal stability. Therefore, the close spatial proximity of two celestial sources would also entail a close temporal coverage by eROSITA and XMM-Newton.

To infer SWCX variation, we focus on the background region count rate during each visit instead of the neutron star itself. MOS2 is a natural choice; all six outer ring CCDs remained operational despite being configured to the small window mode. On the other hand, two of the outer ring CCDs of MOS1 were lost due to micrometeoroid impacts, decreasing the background area. EPIC-pn was configured to the small window mode; therefore, it does not provide sufficient background region for analysis. 

We found the count rates were high at the beginning of all four visits due to the radiation belts. These periods were removed. We note that XMM-Newton observations are affected not just by heliospheric SWCX but also by magnetospheric SWCX. We attempt to minimise the contribution from the latter, at least the time-variable component (variable in a time scale of hours, see \citet{Kuntz2019}) of it, by filtering flares in the softer band ($0.3$--$0.7$\,keV) within the comparatively long ($\gtrsim 20$\,h) exposure time in each visit. We found no flares in the first two visits, and $\sim 8\%$ of exposure times were discarded in the third visit from the filtering.
Unfortunately, the fourth visit on 11 October 2021 suffered from an enhanced quiescent background which cannot be cleaned. Further investigation into the radiation levels from all the instruments on board XMM-Newton reveals that a coronal mass ejection likely caused this. eROSITA also suffered from an enhanced background during this period; however, we could identify and remove this period from the data because of the much more extended coverage. Moreover, CCD5 of MOS2 were in an anomalous state at least during the first and third visits, where the background below 1~keV was strongly enhanced \citep{Kuntz_Snowden08}. To err on the safe side, we removed CCD5 in all visits from the analysis.

From the sight line of CrA, we found the least increase in SWCX from eRASS1--4. Models with a constant SWCX could also fit the eRASS1--4 data, considering the measurement uncertainties. Nonetheless, we calculate the background count rates from the RX~J1856 datasets to inspect the match. We report count rates from two bands --- $0.3$--$0.7$~keV for variation in SWCX and $8$--$12$~keV for instrumental and particle background monitoring. The count rates and the eROSITA and XMM-Newton coverage period, are listed in Table\,\ref{tab:rxj_rate}. 

\begin{table*}[htbp]
    \centering
    \begin{tabular}{cccc}
    \hline \hline
    eROSITA Coverage & XMM-Newton Coverage & $0.3$--$0.7$~keV & $8$--$12$~keV \\ \hline
    2020-Apr-10--2020-Apr-14 & 2020-Mar-31--2020-Apr-01 & $5.989\pm0.045$ & $2.244\pm0.019$ \\
    2020-Oct-13--2020-Oct-17 & 2020-Sep-15--2020-Sep-16 & $6.123\pm0.048$ & $2.286\pm0.020$\\
    2021-Apr-08--2021-Apr-14 & 2021-Apr-01--2021-Apr-02 & $6.101\pm0.046$ & $2.270\pm0.019$\\
    2021-Oct-12--2021-Oct-17 & 2021-Oct-11--2021-Oct-12 & $7.934\pm0.062$ & $2.976\pm0.025$\\
    \hline
    \end{tabular}
    \caption{Background count rate of the field of RX~J1856.5-3754 as observed by XMM-Newton/EPIC-MOS2 in a similar time period of eROSITA's coverage of CrA.}
    \tablefoot{The count rates reported in the $0.3$--$0.7$~keV band are calculated using the vignetting-corrected exposure time, while the $8$--$12$~keV band count rate used the exposure time without vignetting correction, because the latter should be dominated by particles that did not go through the mirror module. They have units of counts\,s$^{-1}$\,deg$^{-2}$.
    The 1\,$\sigma$ error bars are shown.}
    \label{tab:rxj_rate}
\end{table*}

The 0.3--0.7~keV count rates appear to be constant from the period corresponding to eRASS1--3. While the increase from eRASS1 to eRASS2 appears to be $\gtrsim 1\,\sigma$ significance at first glance, but coincidentally the instrumental or particle background was also enhanced as evidenced by the 8--12~keV band, which could likely cause the enhancement. We cannot attribute any count rate increase in eRASS4 to SWCX because of the coronal mass ejection that raised the count rates significantly in both bands (see Sect.\,\ref{sec:data}). In summary, the SWCX variation observed by XMM-Newton in approximately the same epoch as the first three eRASSs seems to agree with eROSITA, which was relatively constant in time towards CrA, despite the uncertainty in determining the magnetospheric SWCX level.
\FloatBarrier

\section{Filter wheel closed data in 020 processing version} \label{sec:FWC}
To study low-intensity diffuse emission, one needs to separate the contribution of the eROSITA instrumental background of the total measured intensity. The most direct way to estimate the instrumental background is by measuring the background level when the filter wheel is rotated to the \texttt{CLOSED} position. For the study of the local hot bubble (LHB) assuming the canonical plasma temperature of $0.1$~keV, the relevant energy range for spectral analysis is $\lesssim 0.6$~keV.
The emission of the LHB is expected to dominate the overall X-ray background at an even lower energy ($< 0.2$~keV) so that spectral analysis in this range would permit more stringent constraints to the properties of the LHB plasma. However, the lack of effective area combined with the rapidly increasing instrumental background of eROSITA $\lesssim 0.2$~keV makes this impossible. Therefore, the useful spectral range to constrain the LHB is limited to $0.2 < E \lesssim 0.6\,\mathrm{keV}$.
The main sources of instrumental background at this energy range include electronic noise arising from the circuitry and secondary X-rays created by high-energy particles hitting the camera \citep{Freyberg20}.
The background induced by high-energy particles, commonly referred to as the particle background, does not dominate in the energy range concerned. The particle background dominates the spectrum at $E > 4$~keV in the case of eROSITA. Due to an update of the pipeline processing of eROSITA data since version 010 (also implemented in the newer version 020), which includes improvements to the pattern identification algorithms to suppress the electronic noise at the low energy end, the instrumental background also changes accordingly. In this section, we present the filter-wheel-closed (FWC) data in the latest 020 version at the time of writing and their extraction procedures.

The 020 processing version provides the broadest uniformly calibrated eROSITA dataset thus far, in the sense that all retrieved eROSITA event lists, from the CalPV phase to the completion of eRASS4 of all 7 TMs have been processed. Despite the pipeline processing improvements, its breadth supersedes the 010 version in terms of temporal coverage where only the data prior to eRASS2 were processed. The temporal coverage in the even earlier 946 processing version was similar to 020. However in contrast to 020, TM4 data were not processed after a micro-meteoroid hit the detector on 23 February 2020 \citep{micrometeoroid}. For this reason, we had fewer signals for spectral analysis when using the 946 dataset. Complete temporal coverage from eRASS1 to eRASS4 of all the TMs (only TM1-4 and 6 were used for analysis due to optical light leak in TM5 and 7 \citep{eROSITA}) allows us to have a better estimation of the contamination introduced by the time-variable SWCX component. 

\begin{figure}
    \centering
    \includegraphics[width=0.45\textwidth]{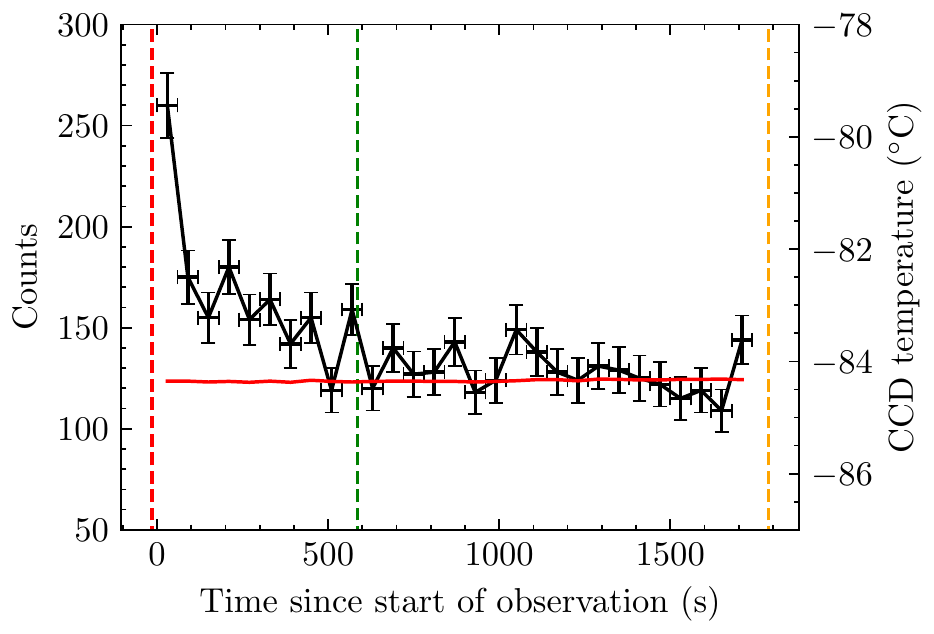}
    \caption{Example FWC light curve in eRASS1 at the $0.5 < E < 10$~keV band. The vertical red dashed line indicates when the filter wheel stopped rotating and stopped at the \texttt{CLOSED} position. The green dashed line indicates the reference cut-off time, which is 10\,min after the red line. The orange line represents the time stamp when the filter wheel rotates again after the observation. Any events recorded between the red and green dashed lines are removed from the FWC dataset, and only events sandwiched between the green and orange dashed lines were considered. The width of the time bin is 60~s. The red horizontal line plots the CCD temperature in \degr C. \label{fig:FWC_lc_eg}}
\end{figure}

Filter-wheel-closed spectra are created by merging all the events when the filter wheel was in the \texttt{CLOSED} position. However, an artefact that plagues most TMs in varying degrees needs to be removed before merging the FWC events, namely the $\sim 10$~min `afterglow' following the rotation of the filter wheel. This afterglow was first seen in ground calibration but was no longer detected after interchanging the materials of the small wheel that drives the filter disc and the inner ring of the filter disc. Unfortunately, the afterglow reappeared after launch, which can be clearly seen in light curves of both FWC and \texttt{FILTER} exposures in all energies above $\sim 0.5$~keV. The afterglow is apparent in an example light curve from TM1 in Figure\,\ref{fig:FWC_lc_eg} as sandwiched by the red and green dashed lines, where the red line indicates the time stamp when the filter wheel rotation ceased and stopped at the \texttt{CLOSED} position and the green line 10\,min after it as reference. The time intervals affected by the afterglow were removed from the good time intervals (GTIs) when extracting the FWC spectra. On top of the afterglow filtering, all FWC observations that have mean count rate deviating from the $90$--$160$~counts\,min$^{-1}$ were removed in order to avoid inclusion of contaminated data\footnote{This filtering criterion also applies to light curves that have CHOPPER $> 1$. In these cases, the bin width increases dynamically to accommodate the changes in CHOPPER. CHOPPER refers to the read-out cadence of the CCD where only $1/\mathrm{CHOPPER}^{\mathrm{th}}$ of all frames would be read out, stored and transmitted back to Earth. CHOPPER $>1$ is used to reduce telemetry when encountering bright sources. Otherwise, in most pointing directions, $\mathrm{CHOPPER=1}$.}. Moreover, a basic flag-filtering was applied to the FWC data before creating the spectra. The flag selection parameter used in the eSASS task \texttt{evtool} is 0xE000F000, which primarily rejects events detected by bad pixels and in the out-of-the-field-of-view CCD area. We note that this choice of flag selection is identical to that applied to all science data processed in 020 to ensure the FWC data is representative of the instrumental background in the astrophysical observations. The resulting live times for all TMs are in the order of 100~ks. Table\,\ref{tab:live_time} lists the precise live times of each TM.

\begin{table}[htbp]
    \centering
    \begin{tabular}{cc}
    \hline \hline
    TM & Live time (s)\\ \hline
    1  & 138257 \\
    2  & 117327 \\ 
    3  & 88255 \\
    4  & 125169 \\
    6  & 106495 \\\hline
    \end{tabular}
    \caption{FWC live times of the TM1--4 and 6.}
    \label{tab:live_time}
\end{table}

\begin{figure*}
    \centering
    \includegraphics[width=0.45\textwidth]{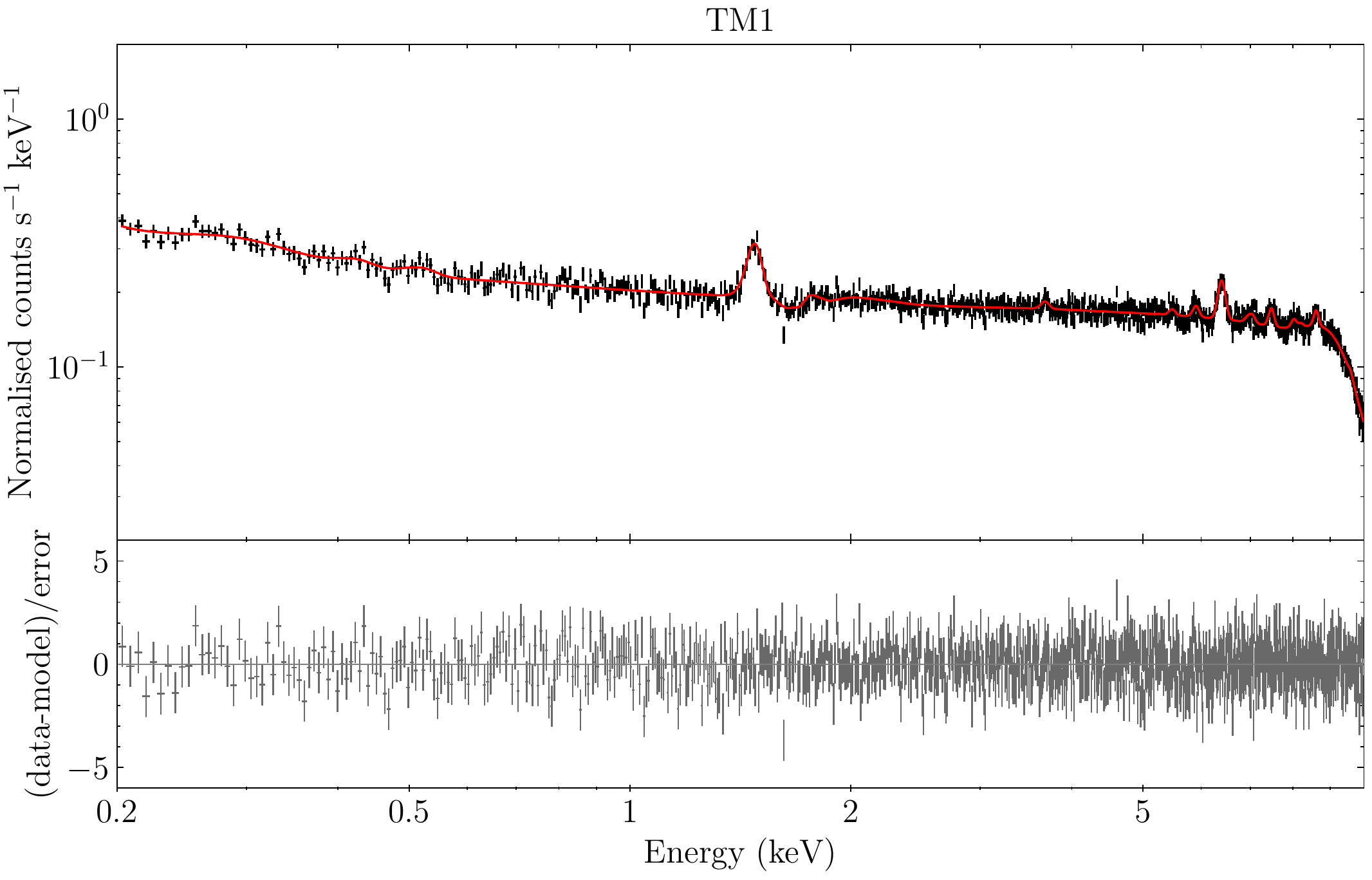}
    \includegraphics[width=0.45\textwidth]{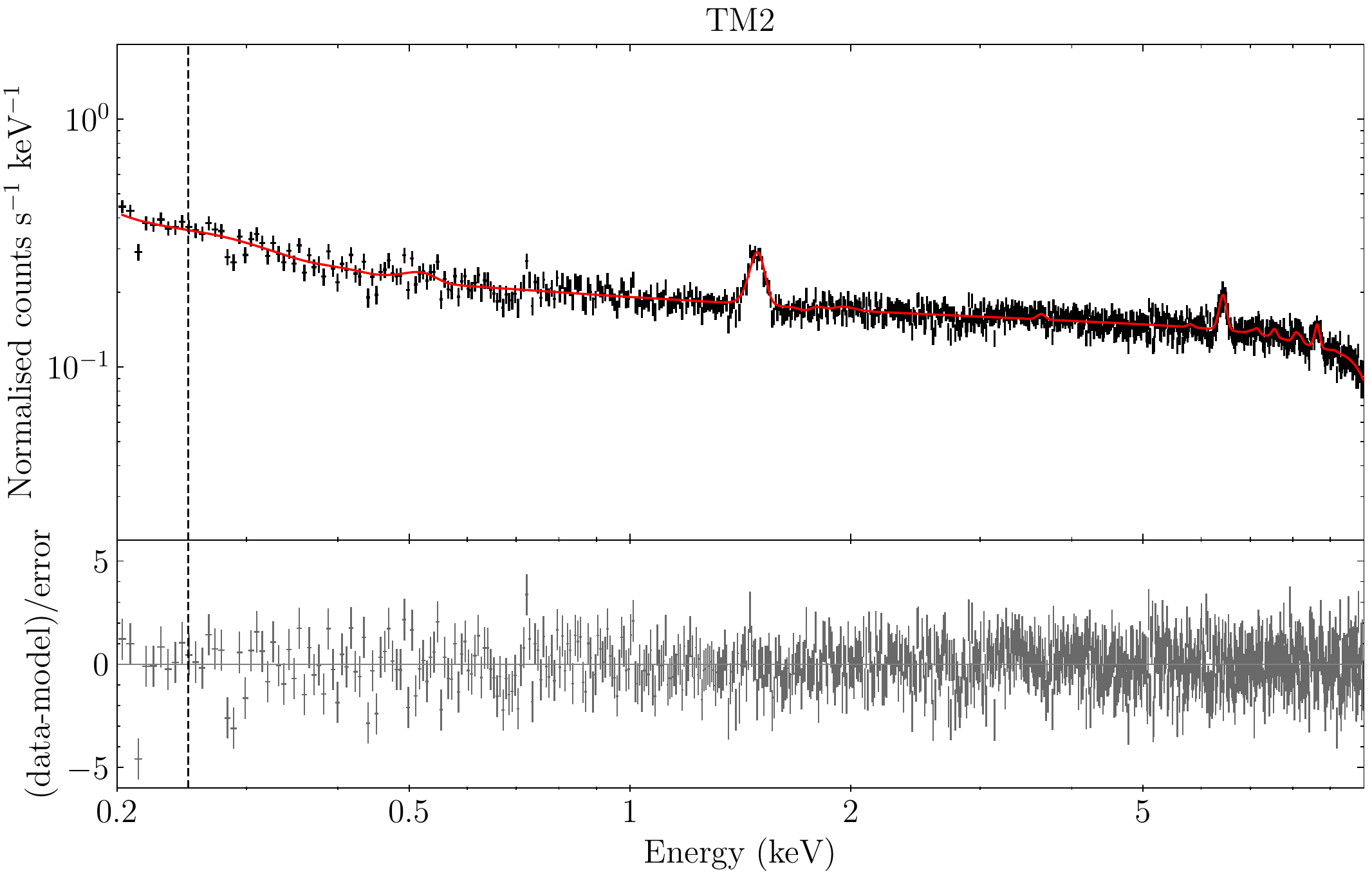}
    \includegraphics[width=0.45\textwidth]{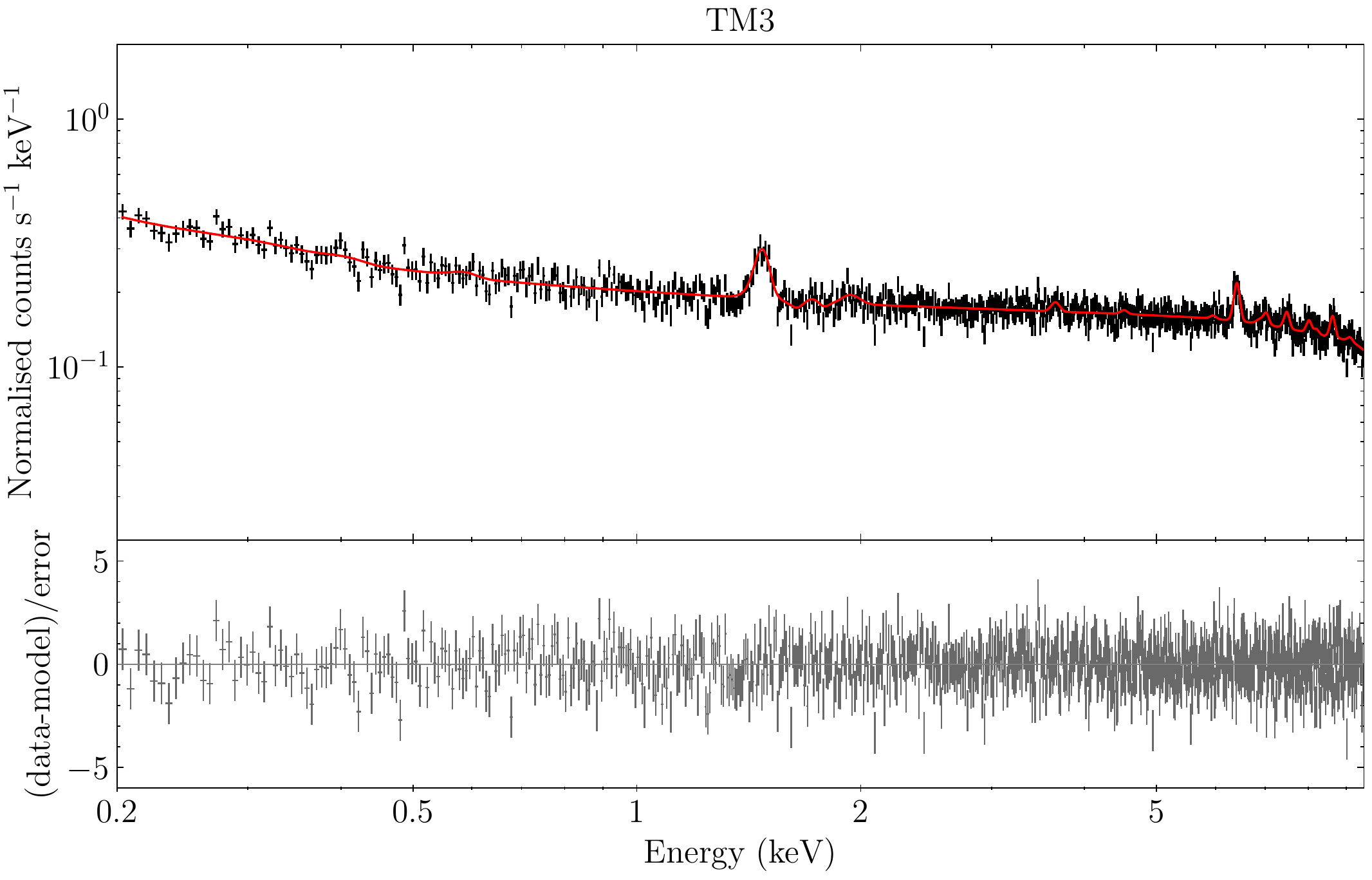}
    \includegraphics[width=0.45\textwidth]{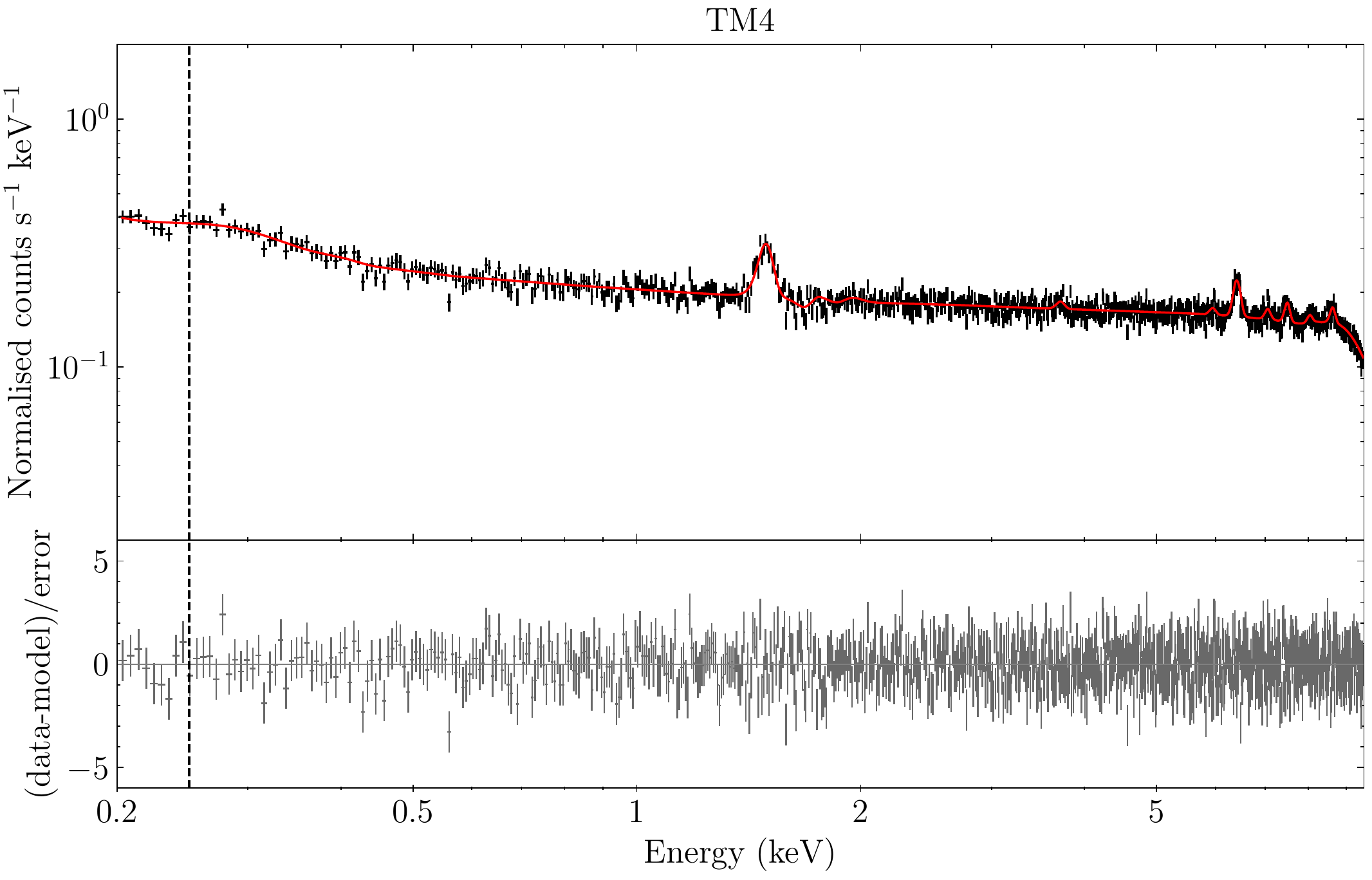}
    \includegraphics[width=0.45\textwidth]{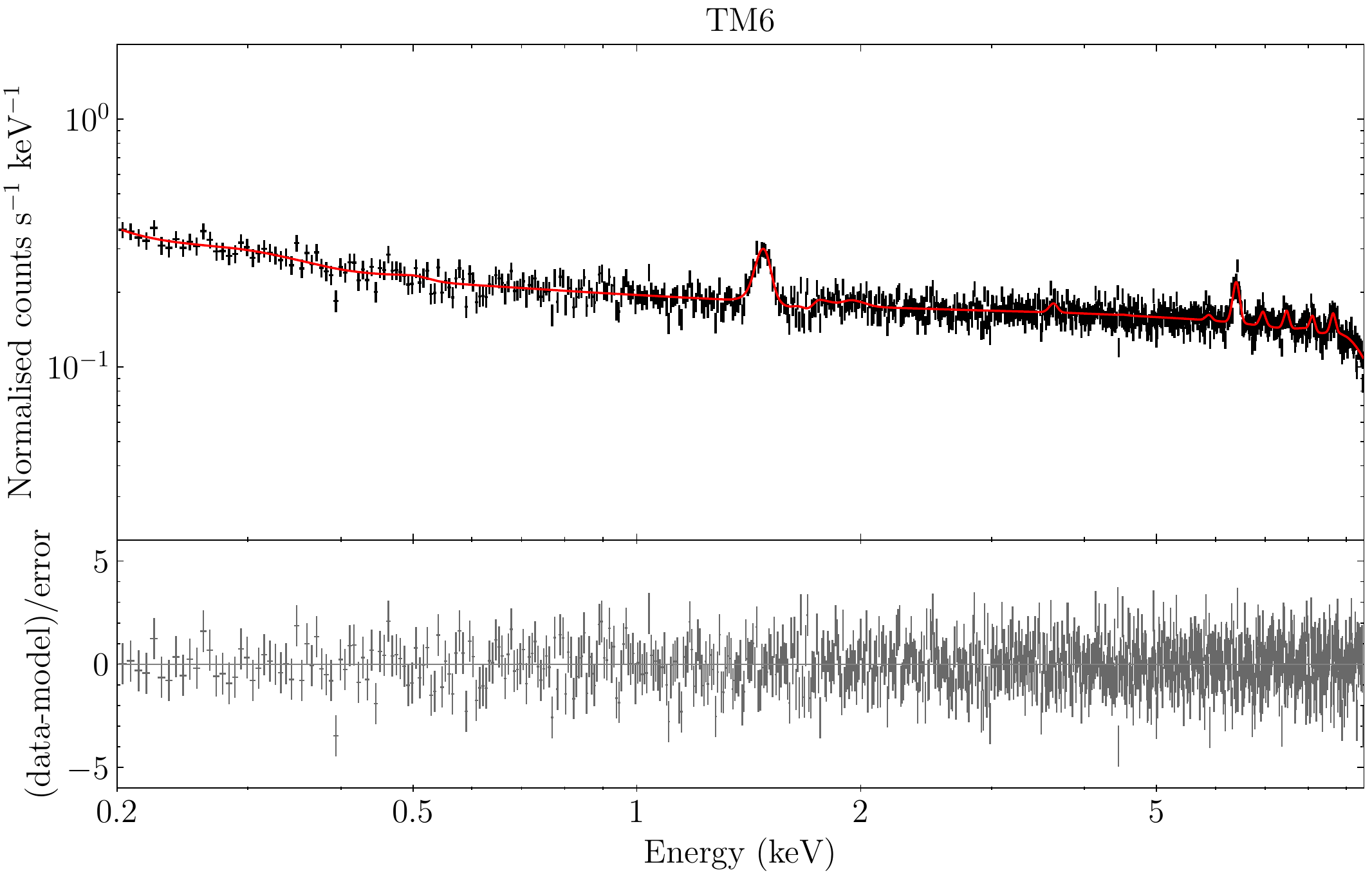}
    \caption{FWC spectra of all patterns for TM1--4 and 6. The best-fit FWC model is shown in red. The vertical dashed lines show the lowest energy the FWC model is employed due to variability considerations (Section~\ref{subsec:FWC_time}).}
    \label{fig:TM1_FWC_pat15}
\end{figure*}

Figure\,\ref{fig:TM1_FWC_pat15} shows the FWC spectra combining all patterns overlaid with the best-fit models in red for TM1--4 and 6. All TMs possess an approximately flat spectrum between $\sim 1$--$9$~keV, then a variable cut-off between $9$--$10$~keV, depending on the TM. The variable cut-off is caused by the minimum ionisation particle threshold being defined on the raw event amplitudes instead of energy. Thus the cut-off varies according to the gain of each TM \citep{Freyberg20}. Fluorescence lines including the most prominent \element{Al} K$\alpha$ at $1.5$~keV and \element{Fe} K$\alpha$ at $6.4$~keV are clearly visible. The flourescence lines are believed to originate within the cameras from the graded shield consisting primarily of \element{Al}, \element{Be}, plus some impurities including \element{Fe} in the Be layer \citep{eROSITA}. A gradual enhancement below $1$~keV is mainly induced by electronic noise. 

The FWC spectral models provide a convenient way to subtract the instrumental background from the spectra of extended sources or diffuse emissions. These models aim to reproduce the FWC data as closely as possible instead of being fully physically motivated. The main components include power-law models to trace the overall shape of the continuum with an exponentially decaying tail $\sim9$--$10$~keV to reproduce the cut-off. Fluorescence lines are reproduced by introducing gaussians at the appropriate energies. In addition to the florescence lines, gaussians are also introduced to modulate the model to accommodate small perturbations deviating from a perfectly smooth continuum.

There is an assumption that the FWC spectra do not change with time and are stable despite variations in CCD temperature to justify our use of a single FWC model for each TM. In the Section~\ref{subsec:FWC_time}, we demonstrate the variabilities of the FWC spectra regarding the time evolution and the CCD temperature. 

\subsection{Variability of FWC spectra} \label{subsec:FWC_time}
\begin{figure*}
    \centering
    \includegraphics[width=0.45\textwidth]{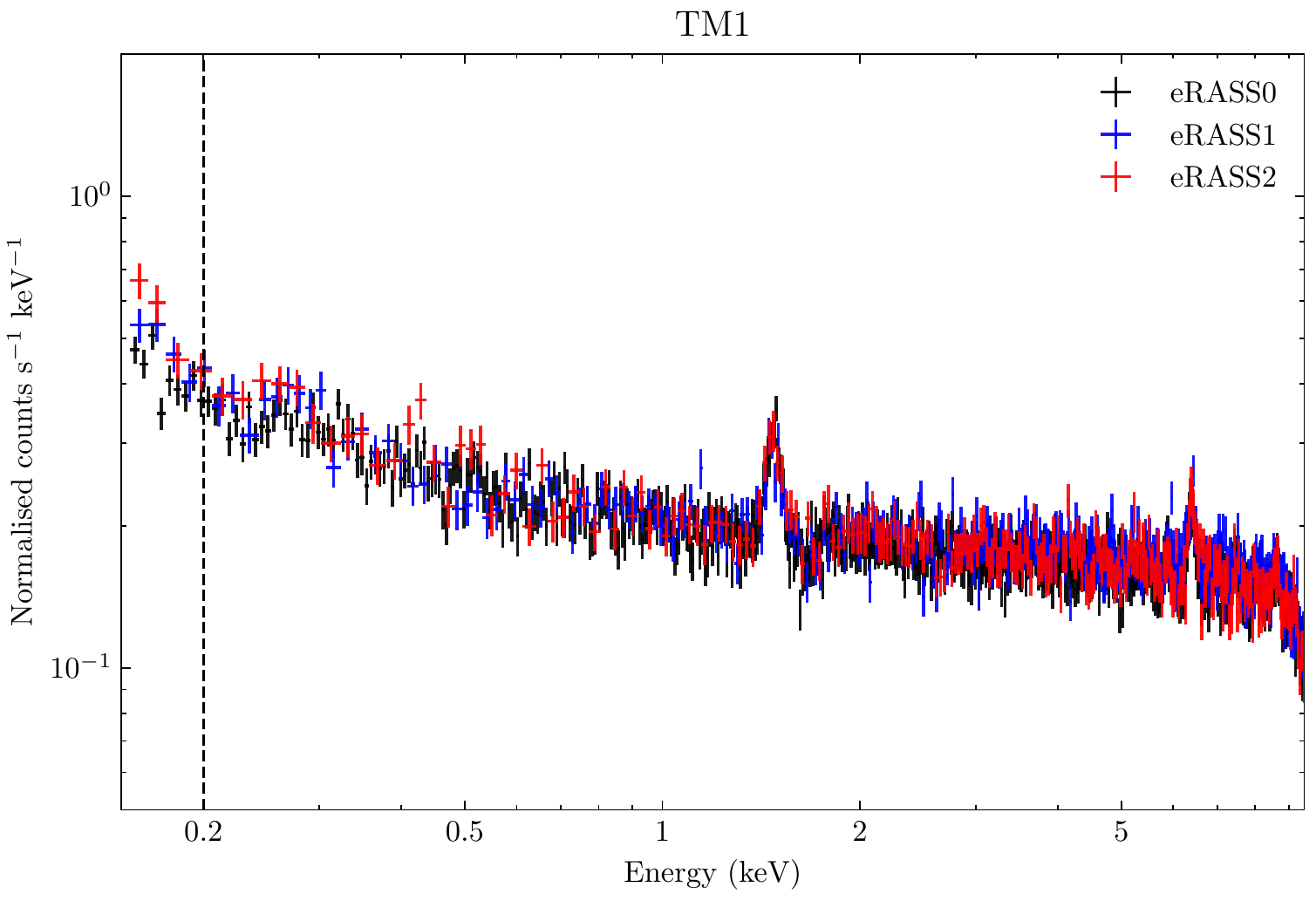}
    \includegraphics[width=0.45\textwidth]{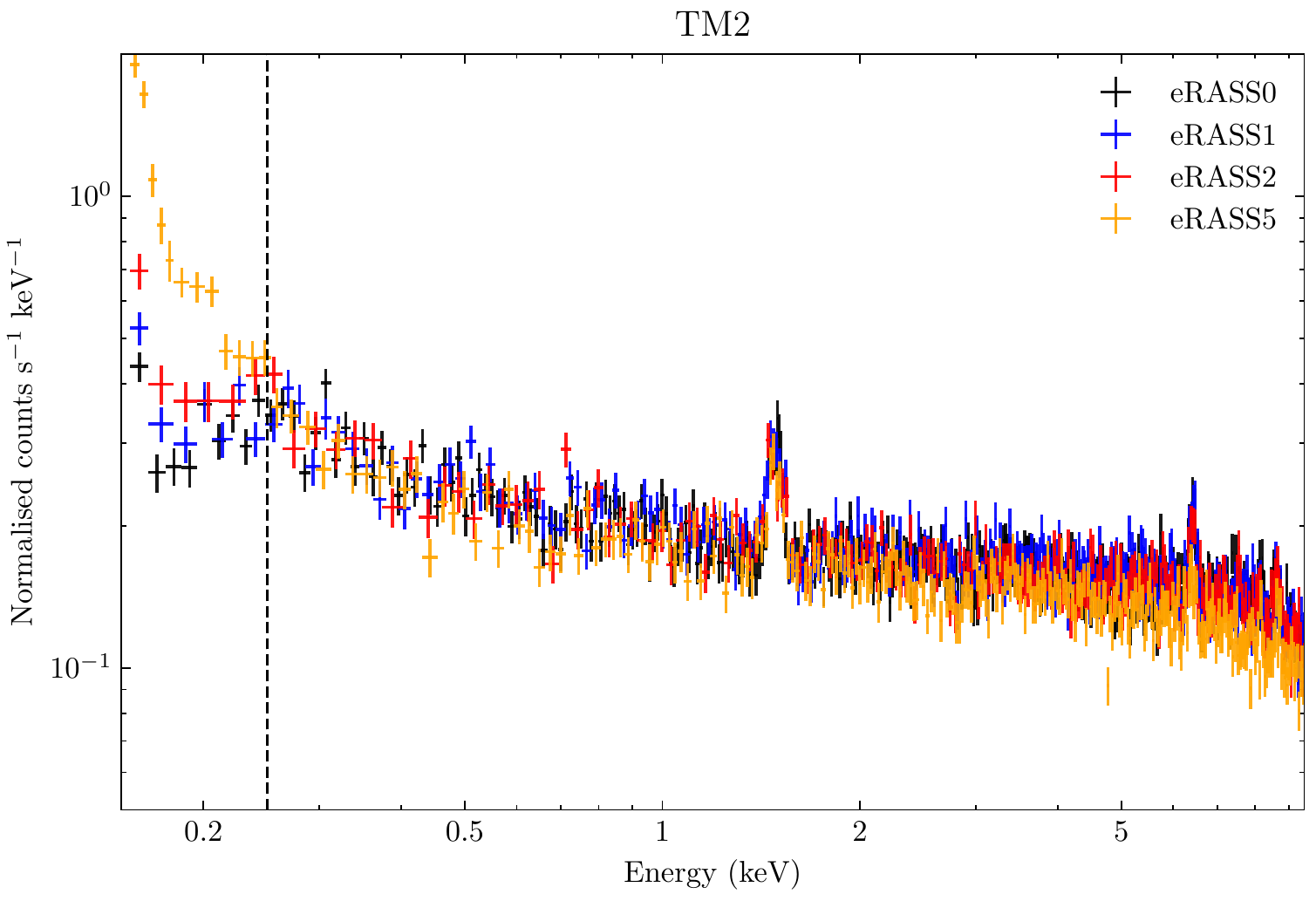}
    \includegraphics[width=0.45\textwidth]{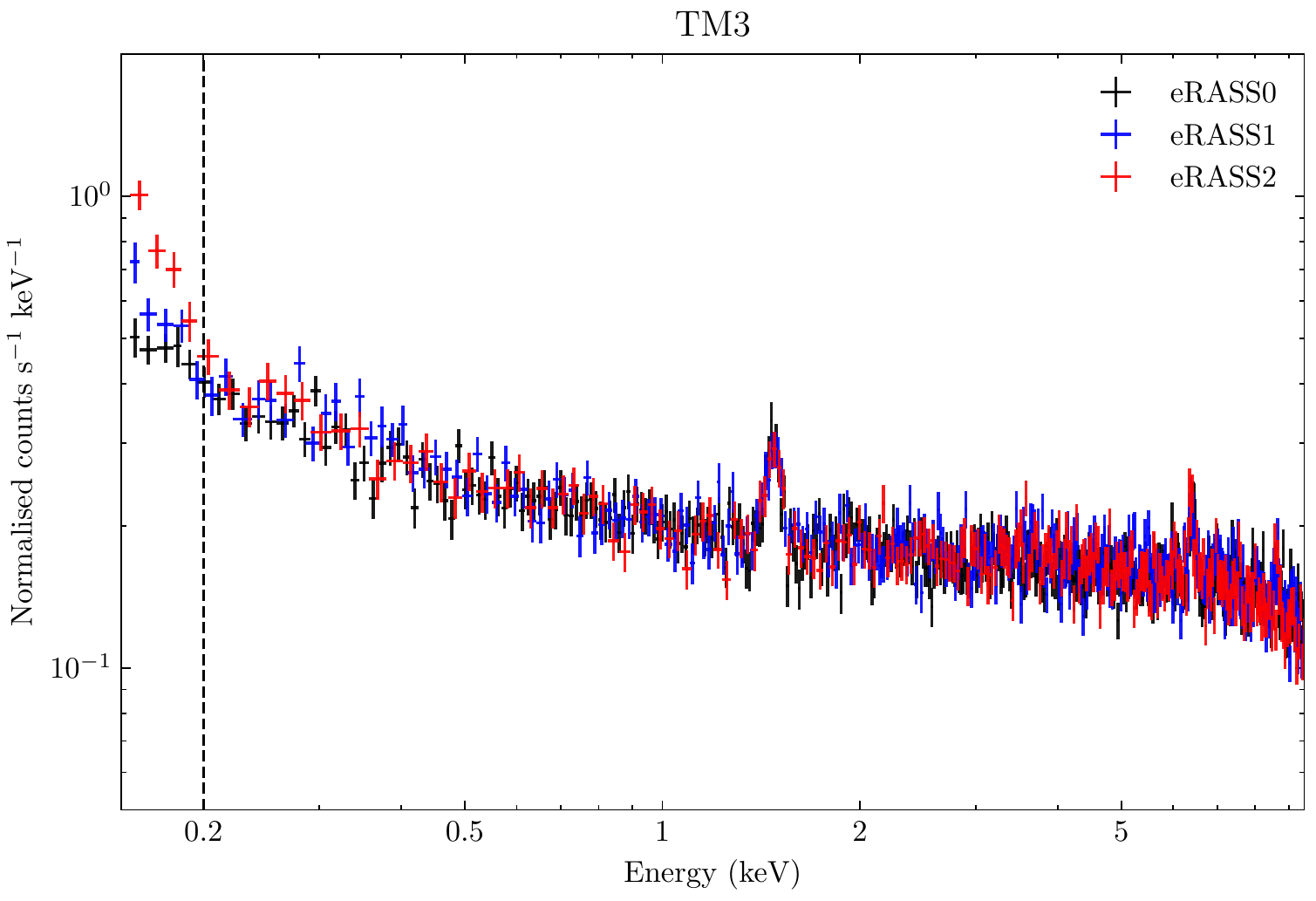}
    \includegraphics[width=0.45\textwidth]{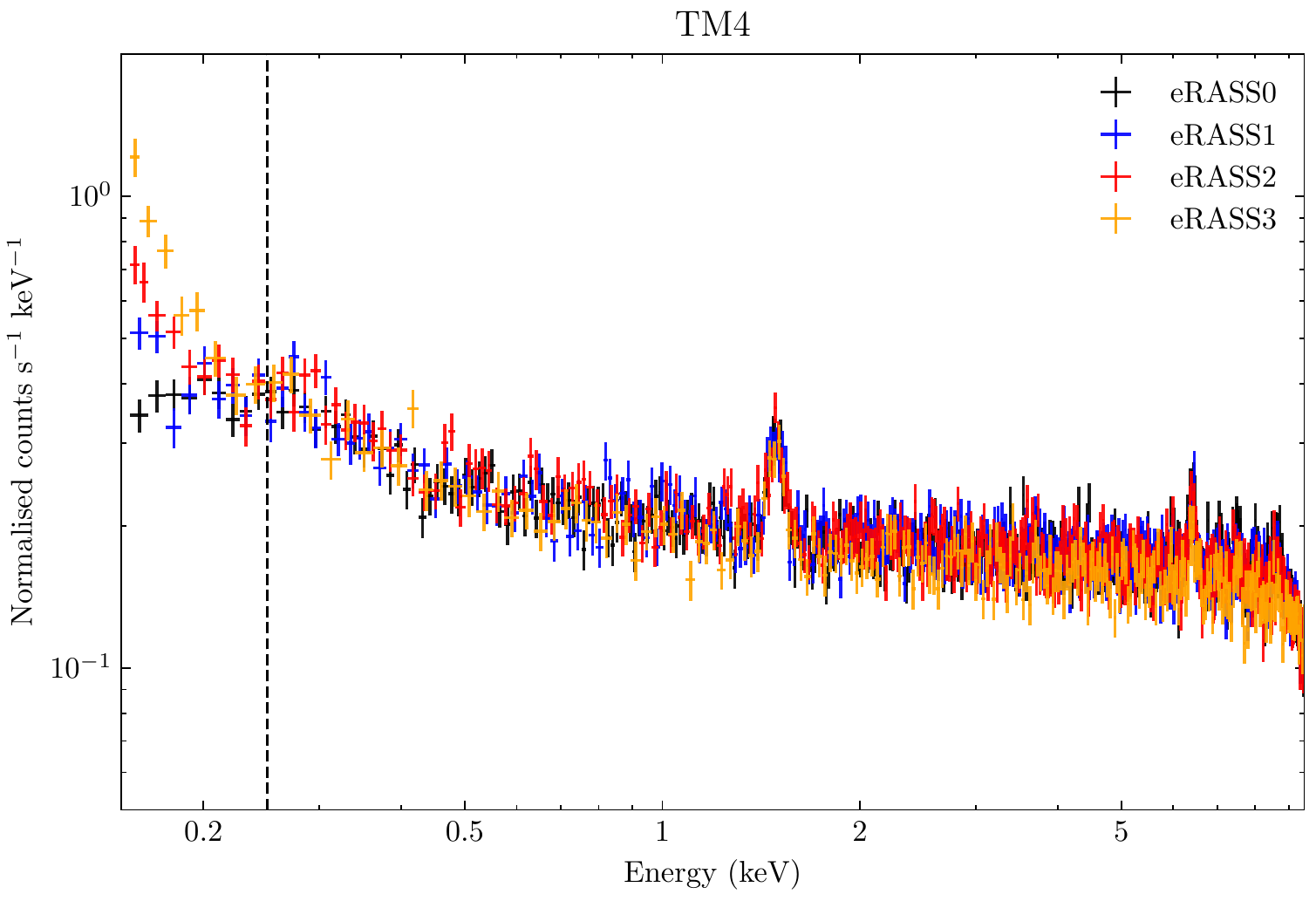}
    \includegraphics[width=0.45\textwidth]{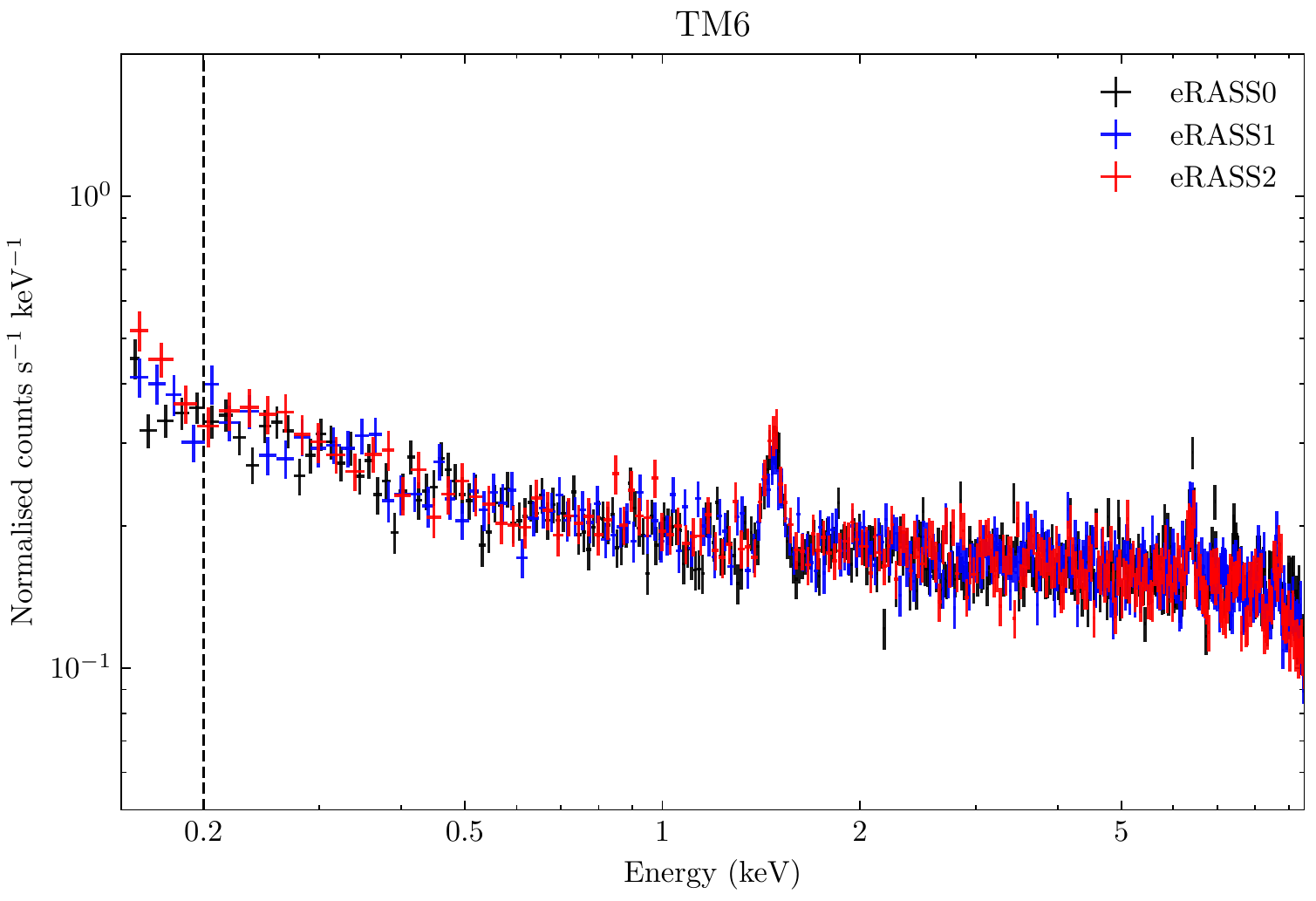}
    \caption{FWC Spectra of each eRASS including the CalPV phase. The vertical dashed lines indicate the energy below which variability sets in.}
    \label{fig:FWC_time_var}
\end{figure*}

Figure~\ref{fig:FWC_time_var} shows the integrated FWC spectra for each eRASS, where eRASS0 corresponds to times in the commissioning, calibration and performance verification (CalPV) programme which preceded eRASS1. FWC observations were conducted for all TMs in eRASS0--2, while only TM4 and TM2 had FWC observations in eRASS3 and 5, respectively. For TM1, 3, 4 and 6, no noticeable variability was observed across $0.2<E<9$~keV. For TM2, the spectrum in eRASS5 is more enhanced $\lesssim0.3$~keV, and slightly below the spectra of earlier eRASSs between $2$--$9$~keV. The lower particle background above $2$~keV is expected due to the well-known anti-correlation of Galactic cosmic ray flux with solar activity (the Sun was approaching maximum activity from eRASS0 to 5), where the solar wind provides some degrees of shielding to Galactic cosmic rays \citep{Neher62,Bulbul20}. Given that the FWC spectrum of TM2 is stable in the $0.25$--$2$~keV range, the energy range for spectral analysis of eRASS3--4 data is limited to $>0.25$~keV. 

For all TMs at $E \lesssim 0.2$~keV, there is a clear trend of higher electronic noise for later eRASSs, despite the exact difference being dependent on the TM concerned. This could be the effect of CCD degradation caused by, for instance, radiation damage.

\begin{figure*}
    \centering
    \includegraphics[width=0.45\textwidth]{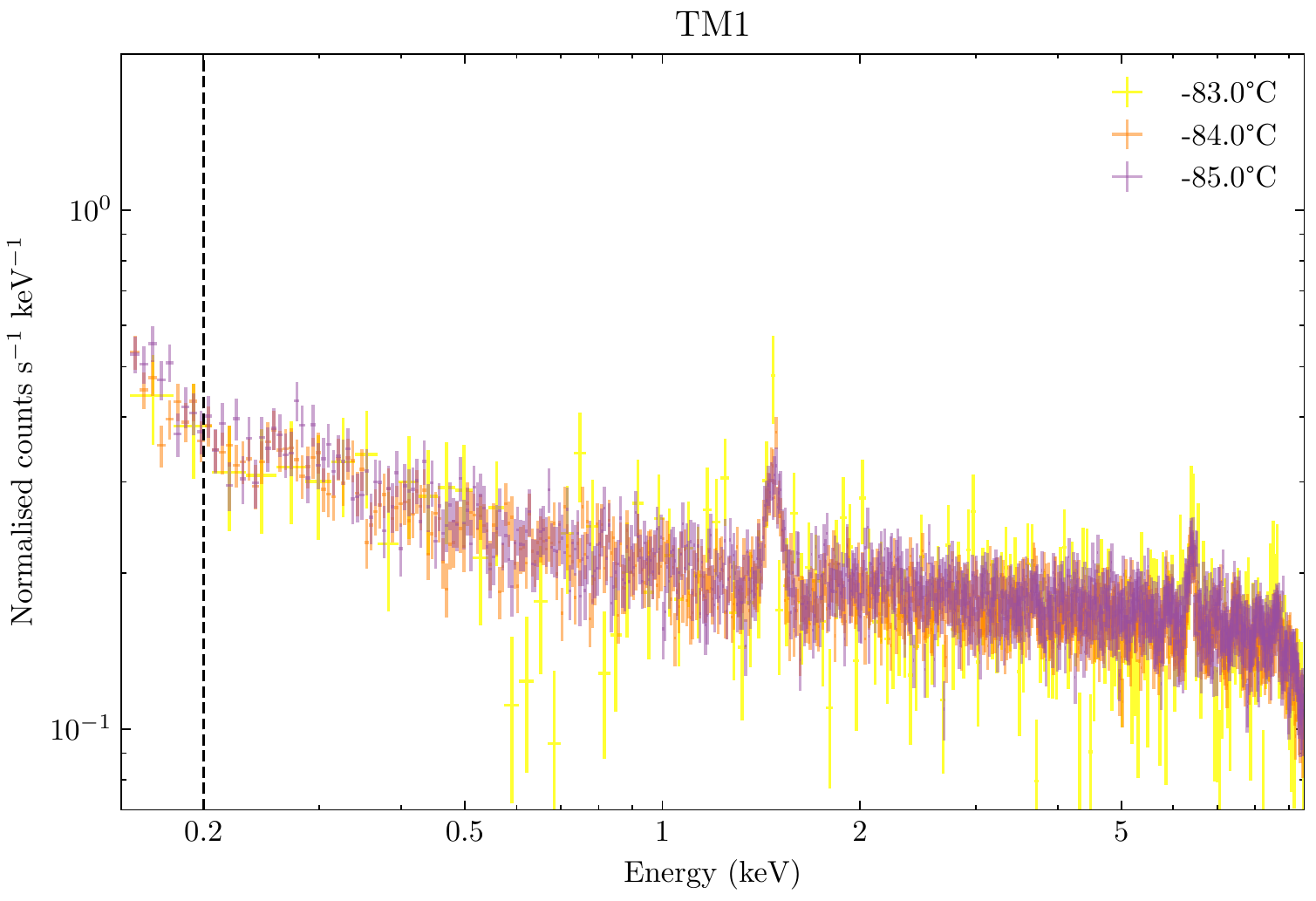}
    \includegraphics[width=0.45\textwidth]{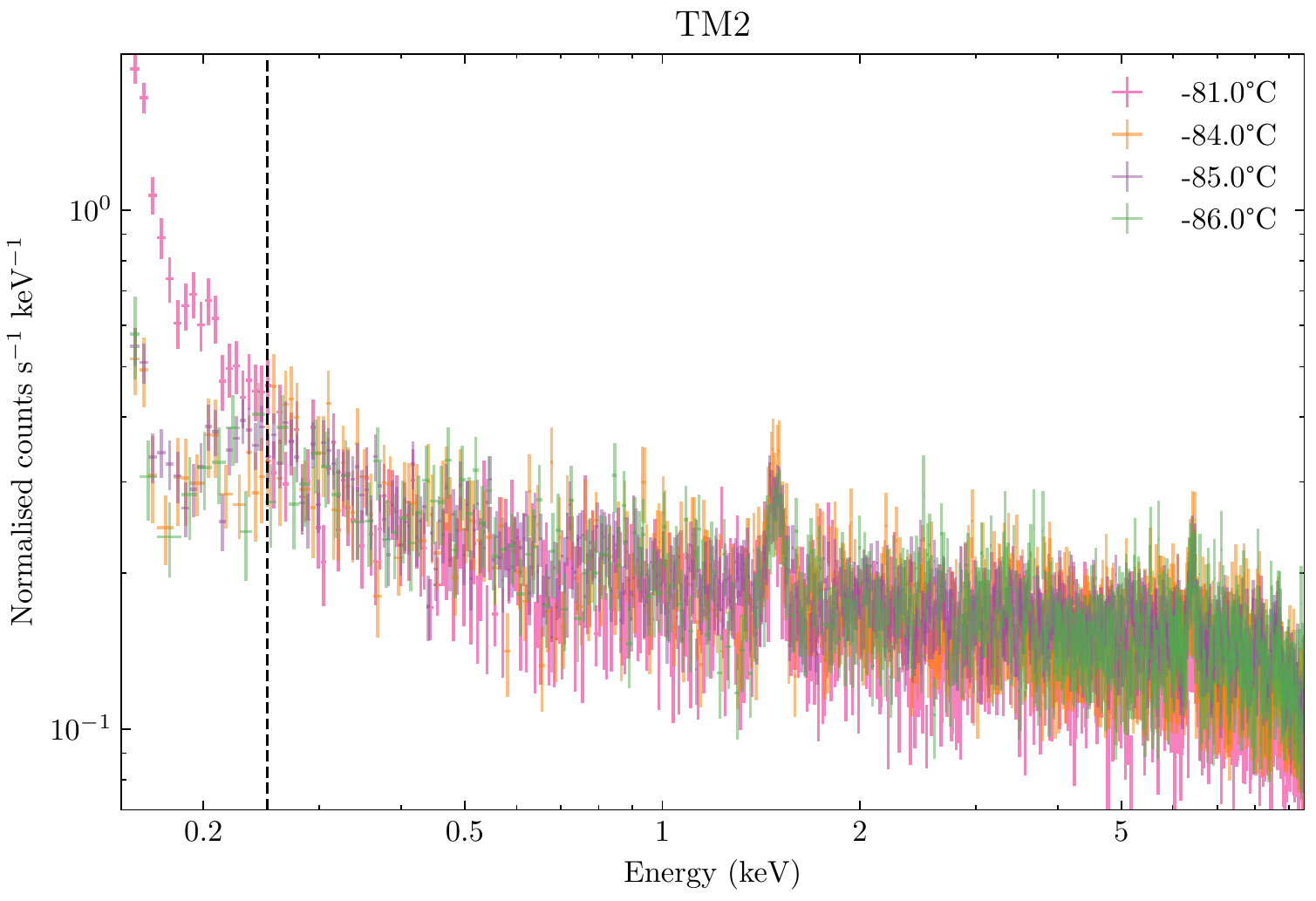}
    \includegraphics[width=0.45\textwidth]{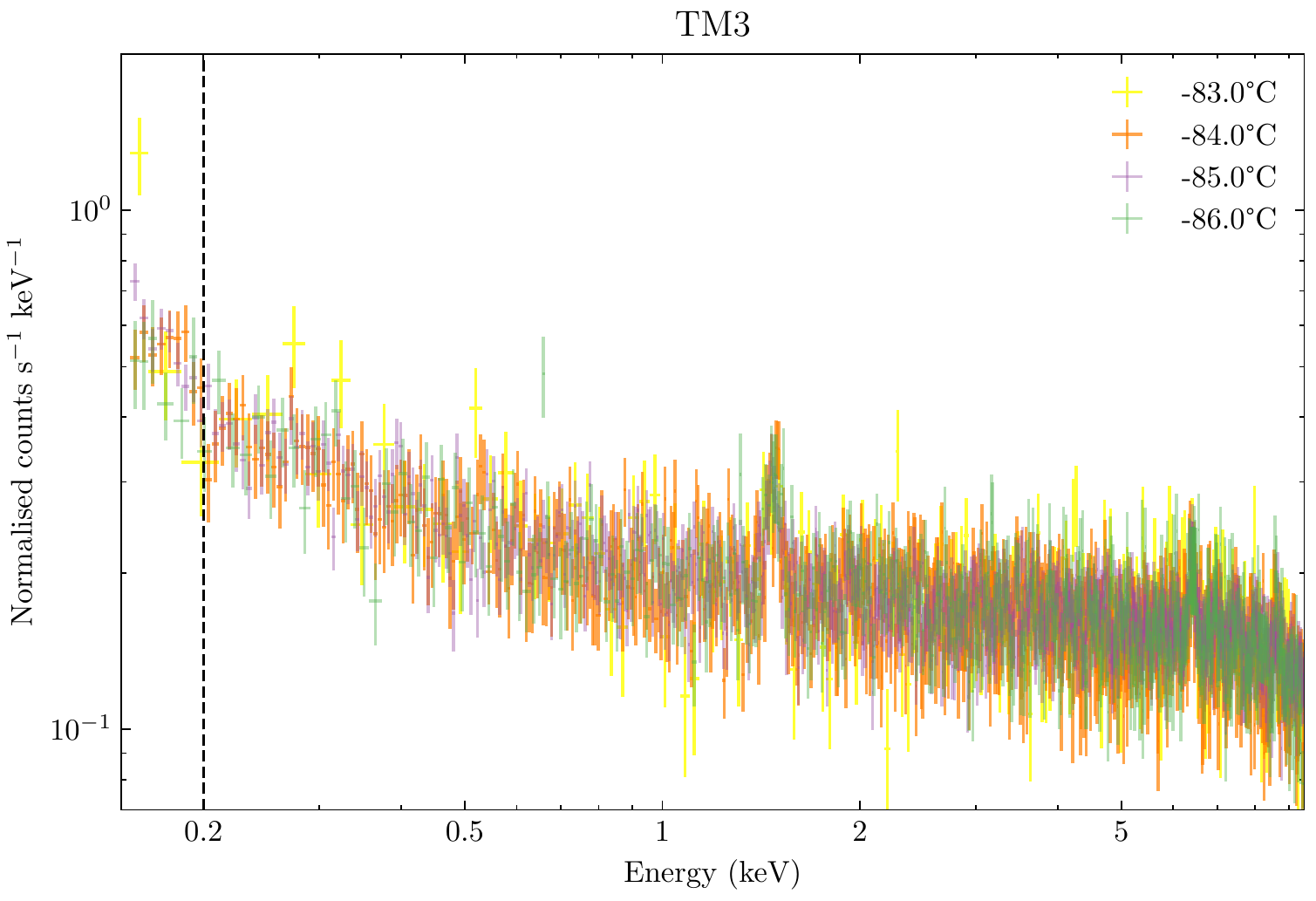}
    \includegraphics[width=0.45\textwidth]{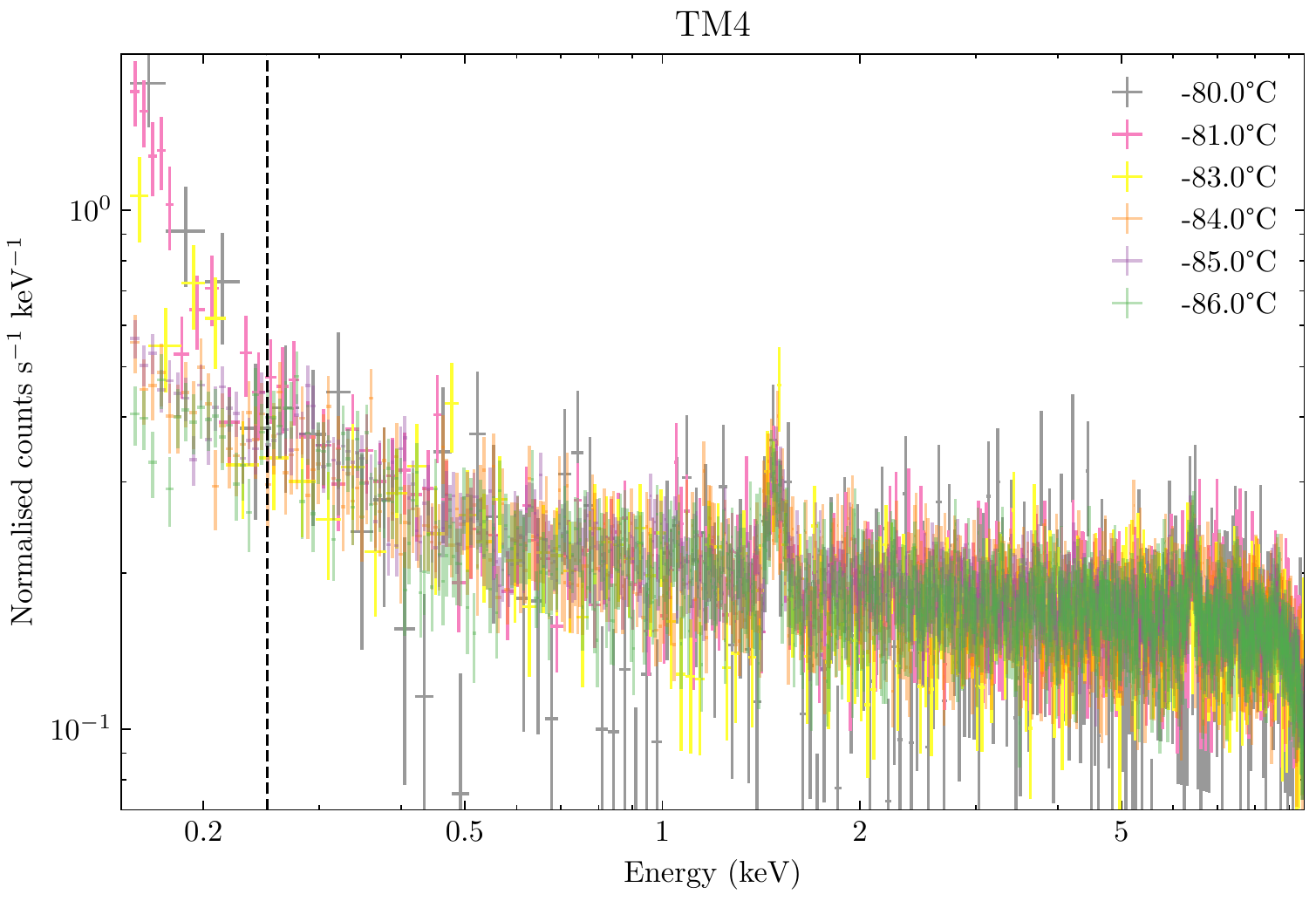}
    \includegraphics[width=0.45\textwidth]{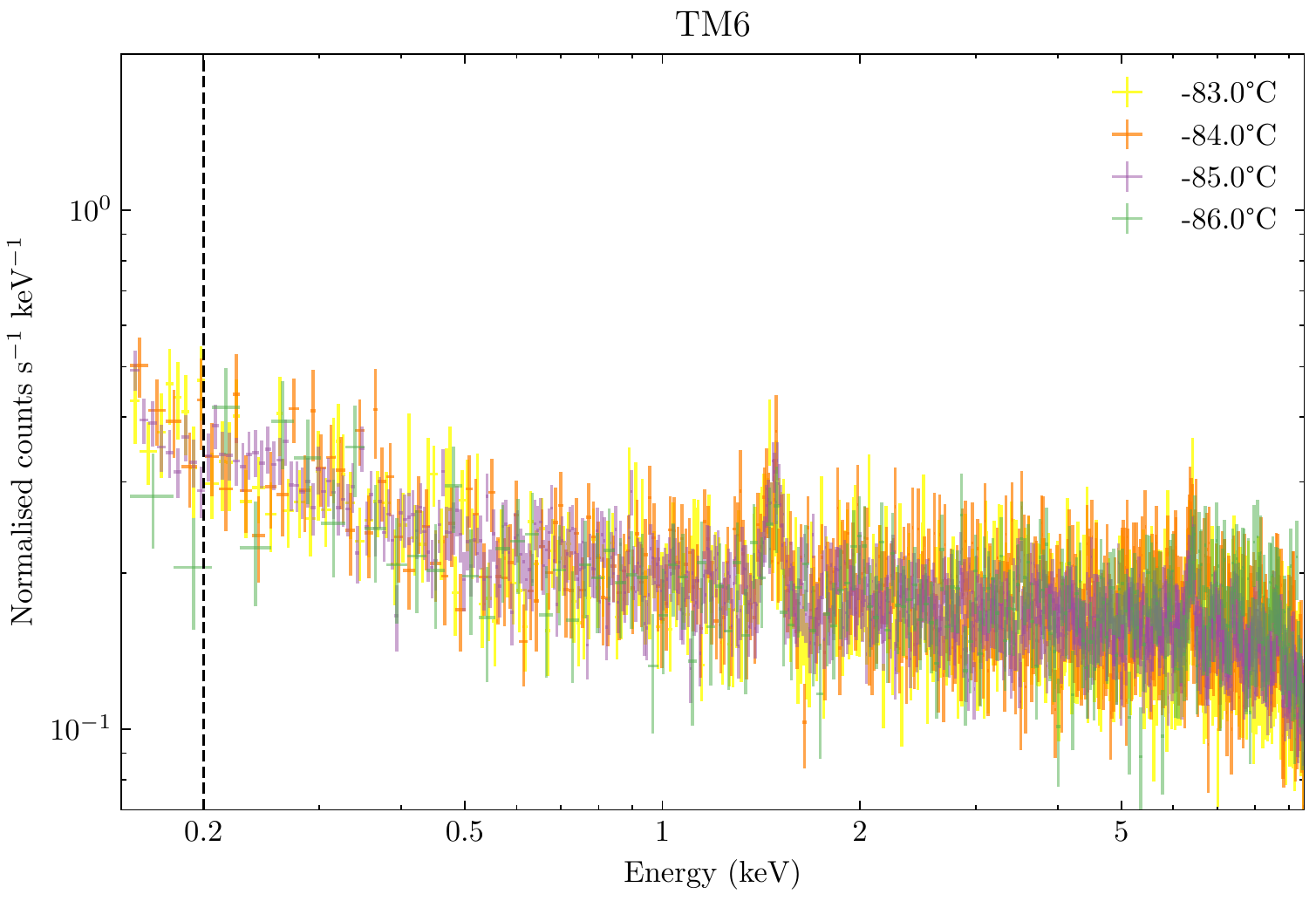}
    \caption{FWC Spectra of as a function of CCD temperature. Each temperature bin has a width of $1$~\degr C centering at the temperature indicated in the legend. The size of the error bars reflect the duration the detector was in the corresponding temperature when the FWC observations were taken. The vertical dashed lines indicate the energy below which variability sets in.}
    \label{fig:FWC_temp_var}
\end{figure*}

CCD temperature is expected to correlate with the electronic noise that contributing to the rising background at low energies. To investigate this effect, we extracted the CCD temperature records from the eROSITA housekeeping files to create GTIs which divide the FWC data into temperature bins of $1$~\degr C width. Figure~\ref{fig:FWC_temp_var} shows the FWC spectrum of each temperature bin. Similarly to time variability, no significant difference between spectra of each temperature is observed for TM1, 3 and 6 above $0.2$~keV. For TM2 and 4, for temperature bins centred at or above $-82$~\degr C, noticeable enhancements $\lesssim0.25$~keV can be observed. While the shapes are similar, the cause of the enhancements is not necessarily identical. Further inspection reveals that all the events in the $-81$~\degr C bin of TM2 are from eRASS5, which means one is essentially looking at the same spectrum shown in Fig.~\ref{fig:FWC_time_var}. Therefore, it is unclear whether the enhancement is a pure result of the degradation of the CCD or it also involves genuine CCD temperature dependence. It is most likely that both effects are present.
On the other hand, despite most of the events in the $-81$ and $-80$~\degr C bins being recorded in eRASS3 in TM4, the eRASS3 spectrum of TM4 (Fig.~\ref{fig:FWC_time_var}) includes events also from lower CCD temperatures. The eRASS3 spectrum between $0.15$ and $0.25$~keV is slightly lower compared to the $-80$~\degr C and $-81$~\degr C spectra suggests the events recorded at lower temperatures have suppressed the enhancement in the eRASS3 spectrum. Hence, at least for TM4, one could see that both CCD degradation and temperature affect the electronic noise component $\lesssim 0.25$~keV. 

For each TM, we have identified the lowest energy that the FWC spectrum is stable, namely, $0.2$~keV for TM1, 3 and 6, and $0.25$~keV for TM2 and 4. These energies were plotted on Fig.~\ref{fig:TM1_FWC_pat15}, \ref{fig:FWC_time_var} and \ref{fig:FWC_temp_var} as reference. Naturally, they serve as the lowest energies to which the corresponding FWC models should be applied. For our purpose, they also set the lowest energies for our spectral analysis.
\FloatBarrier

\section{Position and pointing of SRG/eROSITA with respect to the magnetosphere} \label{sec:orbit}
We also checked the position of SRG/eROSITA with respect to the Sun-Earth-L2 line 
at the times of our shadowing observation periods. Table \ref{tab:orbit_info} gives the distances in geocentric solar ecliptic (GSE) coordinates, in units of Earth radii ($R_\mathrm{E}$). In this right-handed system the $+X$ axis points towards the Sun, with the $+Z$ axis towards the North, $-Y$ in the ecliptic plane in the direction of planetary motion. $YZ = \sqrt{Y^2 + Z^2}$ is the distance from the Sun-Earth-L2 line, the larger the value, the more likely SRG/eROSITA is outside of the magnetosheath. The radius of the bow shock at L2 is usually assumed to be of order $100\,R_{\rm E}$ (for a review on Earth's magnetosphere, see, e.g.\ \citep{Borovsky2018} and references therein). During the four eRASSs the values of $YZ$ ranged between 44 and 139 $R_{\rm E}$, while our targets were observed
mostly at $YZ > 90\,R_{\rm E}$, which is outside the magnetosheath. Enhanced variable background was found more frequently for lower values.

\begin{table*}[htbp]
    \centering
    \begin{tabular}{lcccccc}
    \hline \hline
    eROSITA target ($\lambda,\beta$) & eRASS & Dates & X & Y & Z & YZ \\ \hline
       & 1 & 2020-02-27 -- 2020-03-13 & -234.07 & 111.07 &  1.00 & 111.42 \\
        Cha II \& III $(246\degr,-61\degr)$                                  & 2 & 2020-08-26 -- 2020-09-12 & -239.43 & 102.64 &  7.98 & 103.68 \\
                                     & 3 & 2021-02-11 -- 2021-03-10 & -238.59 &  84.28 & 25.69 &  91.48 \\
                                     & 4 & 2021-08-20 -- 2021-09-06 & -238.50 &  90.68 & 27.46 &  95.84 \\ \hline
                      & 1 & 2020-04-10 -- 2020-04-14 & -185.07 & 91.22 & -58.26 & 108.27 \\
     CrA $(284\degr,-14\degr)$                                & 2 & 2020-10-13 -- 2020-10-17 & -183.05 & 75.48 & -67.20 & 101.11 \\
                                     & 3 & 2021-04-08 -- 2021-04-14 & -185.93 & 66.78 & -70.49 &  97.20 \\
                                     & 4 & 2021-10-12 -- 2021-10-17 & -186.52 & 47.65 & -72.95 &  87.26 \\ \hline
                       & 1 & 2020-03-07 -- 2020-03-15 & -224.82 & 122.12 & -11.05 & 122.70 \\
    Oph $(250\degr,-1\degr)$                                 & 2 & 2020-09-06 -- 2020-09-15 & -228.70 & 114.73 &  -7.48 & 115.12 \\
                                     & 3 & 2021-02-23 -- 2021-03-12 & -230.16 & 101.03 &   7.81 & 102.24 \\
                                     & 4 & 2021-08-31 -- 2021-09-10 & -228.11 & 102.67 &   7.63 & 103.33 \\ \hline    \hline
    \end{tabular}
    \caption{Average position of SRG/eROSITA in geocentric solar ecliptic (GSE) coordinates
    with respect to the Sun-Earth-L2 line 
    at the times of the X-ray shadowing observations (with ecliptic look direction indicated), in units of Earth radii. 
    For details on the orbit see \protect\citep{Freyberg20}.}
    \label{tab:orbit_info}
\end{table*}

At the times of Oph observations, SRG/eROSITA was close to the ecliptic plane (low $Z$ values)  and looking almost in the ecliptic plane ($\beta \sim -1^\circ$). In March (eRASS1 and 3), eROSITA was thus looking through the magnetosphere, while in September (eRASS2 and 4), it was looking away when Ophiuchus was in the field of view.
From observations of the South Ecliptic Pole, which was observed every four hours in Survey mode, we could not find any significant difference when being above or
below the magnetosphere.\\

\FloatBarrier

\section{Posterior distributions of model parameters} \label{appendix:post}
\begin{figure*}[htbp]
    \centering
    \includegraphics[width=\textwidth]{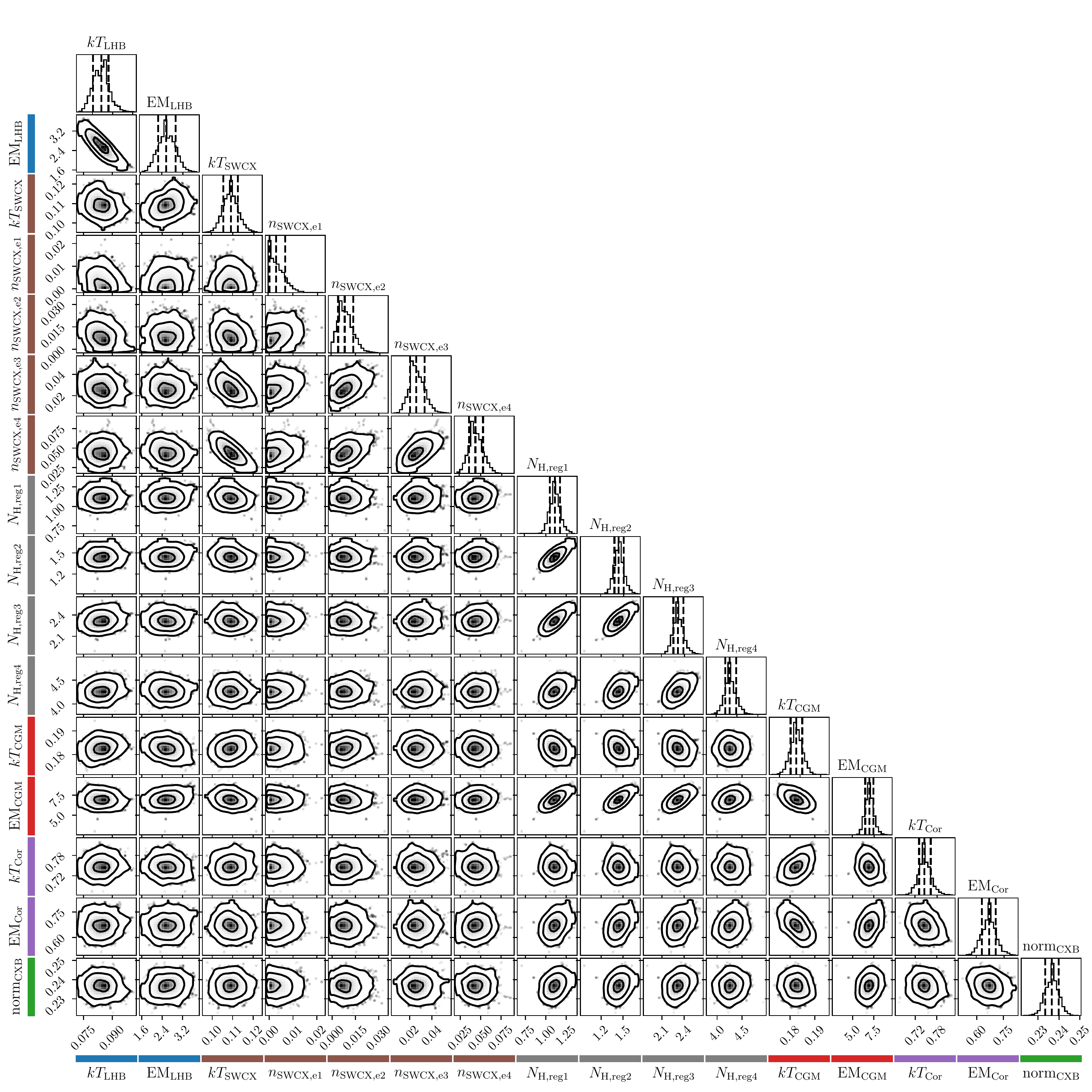}
    \caption{Two-dimensional projections of the posterior distributions of all the model parameters of Cha~II \& III. The contours represent $1$, $2$ and $3\,\sigma$ confidence levels. The parameters are colour-coded using the same colours as in the spectra shown in Fig.\,\ref{fig:cha_fit}, \ref{fig:rcra_fit} and \ref{fig:oph_fit}. The units are the same as in Table\,\ref{table:best-fit}.}
    \label{fig:corner}
\end{figure*}

We present the projected posterior distributions of the model parameters of Cha~II \& III in Fig.\,\ref{fig:corner}. No significant correlations between LHB and SWCX parameters are found. The results for CrA and Oph are similar, thus, are not presented for the sake of brevity. 

\FloatBarrier
\section{Comparison of LHB properties using \texttt{AtomDB} version 3.0.3} \label{sec:apec_v303}

All the spectral fitting results reported in Sect.~\ref{sec:result} are done using the latest \texttt{AtomDB} version 3.0.9. However, to ensure a direct comparison of LHB temperature and EM with the most recent and relevant measurement of the LHB by \citet{Liu16}, we repeat the spectral fitting using \texttt{AtomDB} version 3.0.3. While the \texttt{AtomDB} version adopted by \citet{Liu16} is not specified in their work, version 3.0.3 appears to be a reasonable choice as it would be the latest version six months before its publication, taking into account the reviewing process.

\begin{figure}
    \centering
    \includegraphics[width=0.45\textwidth]{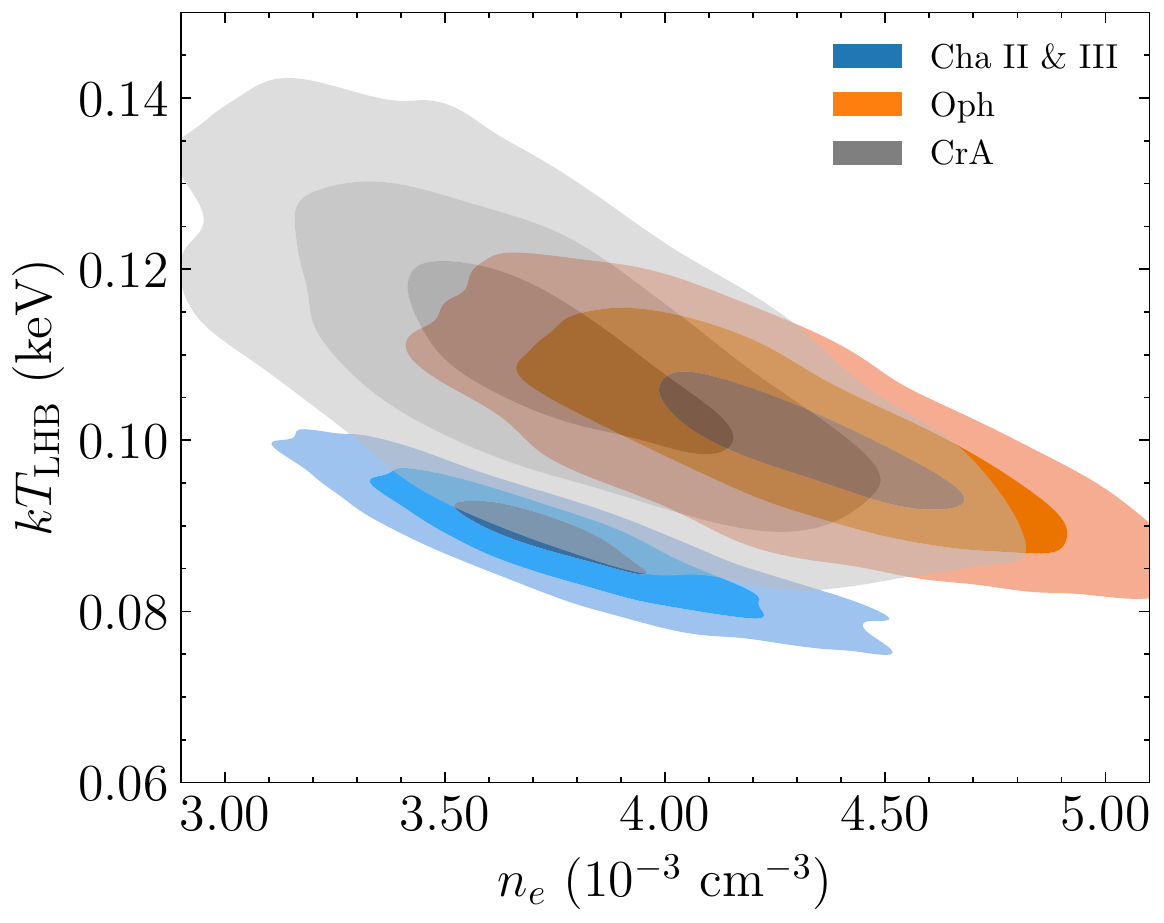}
    \caption{Posterior distributions of the temperature and electron density of the LHB using \texttt{AtomDB} version 3.0.3. The contours indicate the 1, 2 and 3\,$\sigma$ confidence levels (enclose $\sim$\,39, 86 and 99\% of the probability from the highest density).}
    \label{fig:temp_vs_den_apec303}
\end{figure}

We find marginally significant differences towards Cha~II \& III in the foreground components; otherwise, the background components are unaffected. Best-fit parameters of Oph and CrA are not affected by the change of \texttt{AtomDB} version (see Table\,\ref{tab:apec303}).
The increase of $kT_{\mathrm{LHB}}$ of Cha~II \& III from $0.084^{+0.004}_{-0.004}$\,keV (v3.0.9) to $0.089^{+0.003}_{-0.005}$\,keV (v3.0.3) would slightly impact the significance of the LHB temperature difference between the clouds we mentioned in Sect.\,\ref{sec:result}. Using an older version of \texttt{AtomDB} appears to bring $kT_{\mathrm{LHB}}$ measured in Cha~II \& III closer to the other two clouds (Fig.\,\ref{fig:temp_vs_den_apec303}) and the measurement of \citet{Liu16} averaged across the whole sky. Nonetheless, the best-fit $kT_{\mathrm{LHB}}$ and $\mathrm{EM}_{\mathrm{LHB}}$ values of both versions are consistent with the measurement of \citet{Liu16}.

\begin{table}
\caption{Fit parameters of the spectral fitting using \texttt{AtomDB} version 3.0.3.}
\centering
\begin{tabular}{cccc}
\hline\hline
Cloud & Cha\,II \& III & Oph & CrA\\\hline
$kT_{\mathrm{LHB}}$\tablefootmark{(a)} & $0.089^{+0.003}_{-0.005}$ & $0.100^{+0.007}_{-0.007}$ & $0.109^{+0.010}_{-0.010}$ \\ 
$\mathrm{EM}_{\mathrm{LHB}}$\tablefootmark{(b)} & $2.304^{+0.291}_{-0.216}$ & $2.138^{+0.324}_{-0.279}$ & $1.740^{+0.343}_{-0.266}$ \\ 
$kT_{\mathrm{SWCX}}$\tablefootmark{(a)} & $0.108^{+0.003}_{-0.004}$ & $0.109^{+0.002}_{-0.002}$ & $0.105^{+0.004}_{-0.003}$ \\ 
$n_{\mathrm{SWCX, e1}}$\tablefootmark{(c)} & $0.225^{+0.274}_{-0.187}$ & $1.684^{+0.516}_{-0.566}$ & $4.026^{+1.200}_{-1.157}$ \\ 
$n_{\mathrm{SWCX, e2}}$\tablefootmark{(c)} & $0.446^{+0.325}_{-0.273}$ & $4.927^{+0.642}_{-0.629}$ & $4.306^{+1.239}_{-1.116}$ \\ 
$n_{\mathrm{SWCX, e3}}$\tablefootmark{(c)} & $1.566^{+0.437}_{-0.348}$ & $7.277^{+0.782}_{-0.750}$ & $5.623^{+1.239}_{-1.253}$ \\ 
$n_{\mathrm{SWCX, e4}}$\tablefootmark{(c)} & $2.671^{+0.602}_{-0.477}$ & $8.857^{+0.807}_{-0.805}$ & $5.681^{+1.277}_{-1.305}$ \\ 
$N_{\mathrm{H,reg1}}$\tablefootmark{(d)} & $1.154^{+0.069}_{-0.056}$ & $2.410^{+0.196}_{-0.191}$ & $0.644^{+0.042}_{-0.038}$ \\ 
$N_{\mathrm{H,reg2}}$\tablefootmark{(d)} & $1.486^{+0.070}_{-0.064}$ & $3.442^{+0.210}_{-0.204}$ & $1.930^{+0.059}_{-0.044}$ \\ 
$N_{\mathrm{H,reg3}}$\tablefootmark{(d)} & $2.376^{+0.079}_{-0.069}$ & $5.327^{+0.256}_{-0.215}$ & $1.623^{+0.057}_{-0.050}$ \\ 
$N_{\mathrm{H,reg4}}$\tablefootmark{(d)} & $4.348^{+0.122}_{-0.116}$ & $6.377^{+0.417}_{-0.442}$ & $3.670^{+0.086}_{-0.081}$ \\ 
$kT_{\mathrm{CGM}}$\tablefootmark{(a)} & $0.183^{+0.002}_{-0.002}$ & $0.272^{+0.010}_{-0.011}$ & $0.205^{+0.004}_{-0.003}$ \\ 
$\mathrm{EM}_{\mathrm{CGM}}$\tablefootmark{(e)} & $7.260^{+0.538}_{-0.487}$ & $4.435^{+0.812}_{-0.656}$ & $13.674^{+0.719}_{-0.732}$ \\ 
$kT_{\mathrm{Cor}}$\tablefootmark{(a)} & $0.756^{+0.019}_{-0.020}$ & $0.711^{+0.015}_{-0.014}$ & $0.586^{+0.008}_{-0.008}$ \\ 
$\mathrm{EM}_{\mathrm{Cor}}$\tablefootmark{(b)} & $0.661^{+0.045}_{-0.040}$ & $2.371^{+0.177}_{-0.157}$ & $3.383^{+0.131}_{-0.134}$ \\ 
$\mathrm{norm}_{\mathrm{CXB}}$\tablefootmark{(f)} & $0.236^{+0.003}_{-0.003}$ & $0.285^{+0.007}_{-0.006}$ & $0.224^{+0.006}_{-0.006}$ \\ 
\hline
\end{tabular}
\label{tab:apec303}
\tablefoot{The values reported are the $50$ percentiles, with the lower and upper bounds showing the $16$ and $84$ percentiles of the Markov Chain Monte Carlo analysis result.\\
\tablefoottext{a}{$kT_{\mathrm{LHB}}$, $kT_{\mathrm{SWCX}}$, $kT_{\mathrm{CGM}}$ and $kT_{\mathrm{Cor}}$ are in units of keV.} \\
\tablefoottext{b}{$\mathrm{EM}_{\mathrm{LHB}}$ and $\mathrm{EM}_{\mathrm{Cor}}$ are in units of $10^{-3}\,{\rm cm^{-6}\,pc}$.}\\
\tablefoottext{c}{$n_{\mathrm{SWCX}}$ is in the unit of $10^{-2}\,\mathrm{deg}^{-2}$. The normalisation parameter of the \texttt{ACX2} model is dimensionless and is only intended for relative scaling (see the documentation of the \texttt{ACX} model). We normalised this factor by the sky area to give the unit deg$^{-2}$.}\\
\tablefoottext{d}{$N_{\mathrm{H}}$ values are in units of $10^{21}\,\mathrm{cm^{-2}}$.}\\
\tablefoottext{e}{$\mathrm{EM}_{\mathrm{CGM}}$ is in $10^{-2}\,{\rm cm^{-6}\,pc}$.}\\
\tablefoottext{f}{$\mathrm{norm}_{\mathrm{CXB}}$ has unit of $10^{-2}$\,photons\,keV$^{-1}$\,cm$^{-2}$\,s$^{-1}$\,deg$^{-2}$ at 1 keV.} 
}
\end{table}

\FloatBarrier
\end{appendix}

%
%

\end{document}